\documentclass[twocolumn]{aastex62}
\usepackage{amsmath}
\usepackage{color}
\usepackage{natbib}
\usepackage{enumitem}
\usepackage{float}
\usepackage{hyperref}
\usepackage{listings}
\usepackage{xcolor}
\usepackage{xspace} 
\usepackage{url}
\usepackage{fontawesome}
\usepackage{subfigure}

\definecolor{linkcolor}{rgb}{0.,0.3,0.7}
\definecolor{codegreen}{rgb}{0,0.6,0}
\definecolor{codegray}{rgb}{0.5,0.5,0.5}
\definecolor{codepurple}{rgb}{0.58,0,0.82}
\definecolor{backcolour}{rgb}{0.95,0.95,0.92}

\hypersetup{breaklinks, colorlinks,
  urlcolor=linkcolor,linkcolor=linkcolor,citecolor=linkcolor}
\newcommand{\satgen}{\texttt{SatGen}\xspace}

\newcommand{\sersic}{S\'ersic\xspace}

\submitjournal{ApJ}

\graphicspath{{figures/}}

\shorttitle{The Satellite Stellar-to-Halo Mass Relation}

\begin{document}
\title{ELVES IV: The Satellite Stellar-to-Halo Mass Relation Beyond the Milky-Way}

\author[0000-0002-1841-2252]{Shany Danieli}
\altaffiliation{NASA Hubble Fellow}
\affiliation{Department of Astrophysical Sciences, 4 Ivy Lane, Princeton University, Princeton, NJ 08544}
\author[0000-0002-5612-3427]{Jenny E. Greene}
\affiliation{Department of Astrophysical Sciences, 4 Ivy Lane, Princeton University, Princeton, NJ 08544}
\author[0000-0002-5382-2898]{Scott Carlsten}
\affiliation{Department of Astrophysical Sciences, 4 Ivy Lane, Princeton University, Princeton, NJ 08544}
\author[0000-0001-6115-0633]{Fangzhou Jiang}
\affiliation{TAPIR, California Institute of Technology, Pasadena, CA 91125, USA}
\affiliation{Carnegie Observatories, 813 Santa Barbara Street, Pasadena, CA 91101, USA}
\author[0000-0002-1691-8217]{Rachael Beaton}
\affiliation{Space Telescope Science Institute, Baltimore, MD, 21218, USA}
\affiliation{Department of Astrophysical Sciences, 4 Ivy Lane, Princeton University, Princeton, NJ 08544}
\author[0000-0003-4700-663X]{Andy D. Goulding}
\affiliation{Department of Astrophysical Sciences, 4 Ivy Lane, Princeton University, Princeton, NJ 08544}

\begin{abstract}

Quantifying the connection between galaxies and their host dark matter halos has been key for testing cosmological models on various scales. Below $M_\star \sim 10^9\,M_\odot$, such studies have primarily relied on the satellite galaxy population orbiting the Milky Way. Here we present new constraints on the connection between satellite galaxies and their host dark matter subhalos using the largest sample of satellite galaxies in the Local Volume ($D \lesssim 12\,\mathrm{Mpc}$) to date. We use $250$ confirmed and $71$ candidate dwarf satellites around 27 Milky Way (MW)-like hosts from the Exploration of Local VolumE Satellites (ELVES) Survey and use the semi-analytical \satgen model for predicting the population of dark matter subhalos expected in the same volume. Through a Bayesian model comparison of the observed and the forward-modeled satellite stellar mass functions (SSMF), we infer the satellite stellar-to-halo mass relation. We find that the observed SSMF is best reproduced when subhalos at the low mass end are populated by a relation of the form $M_\star \propto M^\alpha_\mathrm{peak}$, with a moderate slope of $\alpha_\mathrm{const}=2.10 \pm 0.01$ and a low scatter, constant as a function of the peak halo mass, of $\sigma_\mathrm{const}=0.06^{+0.07}_{-0.05}$. A model with a steeper slope ($\alpha_\mathrm{grow}=2.39 \pm 0.06$) and a scatter that grows with decreasing $M_\mathrm{peak}$ is also consistent with the observed SSMF but is not required. Our new model for the satellite-subhalo connection, based on hundreds of Local Volume satellite galaxies, is in line with what was previously derived using only the Milky Way satellites.

\end{abstract}

\keywords{methods: observational, methods: statistical, galaxies: dwarf, galaxies: statistics, galaxies: halos, cosmology: observations, cosmology:dark matter}

\section{Introduction} \label{sec:intro}

In our quest to uncover the nature of dark matter, understanding the role of baryons in shaping cosmological structures on small scales has been vital. Dark matter halos embedded within the cosmic web of the large-scale structure are thought to be well-traced by galaxies and gas \citep{Springel:2006, Peacock:2001, Zehavi:2011}. The ever-increasing galaxy samples from galaxy redshift surveys, as well as advances in theoretical models of cosmology and galaxy formation, have led to the formulation of a relationship between galaxies and their host dark matter halos. Many approaches have been used to study this so-called ``galaxy-halo connection'', including various empirical and physical models, as well as utilizing data to constrain the unknown aspects of this relation  \citep[see][for a thorough review]{WechslerTinker:2018}.

One such popular approach is the abundance matching technique, which relies on the premise that the most massive galaxy lives in the most massive halo; the second most massive galaxy lives in the next most massive halo, etc. \citep{KravtsovKlypin:1999}. This assumption is made at a variety of scales, from galaxy clusters to far smaller groups \citep{Maccio:2020}. When considering dark matter subhalos that orbit a group or cluster central potential well, each subhalo within the virial radius of the host halo, massive enough to allow molecular cooling, should in theory host a luminous galaxy \citep{ValeOstriker:2004, Conroy:2006, Guo:2010}. While this basic approach is generally reasonable, when linking galaxies to their host halos, the impact of a number of physical processes should be carefully considered. For example, the strong dynamical evolution of subhalos as they orbit the central galaxy and tidal stripping of subhalos will alter the subhalo mass function, a key ingredient in this method \citep{vdb:2018a, Green:2019}. A more recent implementation of the subhalo abundance matching technique, namely parameterized abundance matching, allows for scatter in the properties of galaxies when these are associated with subhalos of a given mass, which is both physically and observationally motivated \citep{Behroozi:2010, Moster:2010, Matthee:2017}.

The outcome of this procedure is a relationship between the stellar mass and the halo mass of galaxies. This so-called stellar-to-halo mass relation (SHMR) has been extensively studied for central galaxies since the beginning of the 2000s using a variety of methodologies \citep[see][for a complete review on the methods used to model the SHMR]{WechslerTinker:2018}. A key result that emerged from all studies of the SHMR for galaxies above $M_\mathrm{halo} \gtrsim 10^{12}\,M_\odot$ is the remarkably small scatter of $0.2\,\mathrm{dex}$ in galaxy stellar mass at fixed halo mass. \citet{Gu:2016} showed that the small scatter originates in the hierarchical assembly at high masses ($M_\mathrm{vir}>10^{14}\,M_\odot$) and largely by in-situ growth for lower mass halos ($10^{12} < M_\mathrm{vir} < 10^{14.75}\,M_\odot$). This small scatter suggests that galaxy formation is well-regulated at these high masses.

However, below $M_\mathrm{halo} \sim 10^{11}-10^{12}\,M_\odot$, the SHMR is not well-understood for three main reasons. First, present theoretical models still struggle to accurately capture the whole set of physical processes that drive the formation and evolution of low-mass galaxies in their host halos, as well as their relative importance. The impact of feedback from massive stars \citep{Hopkins:2011} and the suppression of star formation caused by reionization \citep{Dawoodbhoy:2018} are a few examples. Secondly, $N$-body simulations used to construct the halo mass function are often close to their resolution limit when low-mass galaxies are considered \citep{vdb:ogiya:2018}. Finally, the observational data employed in the SHMR modeling below $M_\mathrm{vir} \sim 10^{11}\,M_\odot$, involving the dwarf galaxy regime, is observationally far less robust and comprehensive than in the case of their high mass counterparts. 

Nonetheless, significant progress has been made on all three fronts in the last few decades. Modern wide-field imaging surveys have been covering larger areas of the sky, reaching depths that allow the detection and characterization of MW ultra-faint dwarf galaxies with an absolute $V$-band magnitude as low as $M_V \sim -1$ to $-4\,\mathrm{mag}$. The census of dwarf galaxies in the Local Group has increased by an order of magnitude owing to discoveries from the Sloan Digital Sky Survey (SDSS; \citealt{Willman:2005a, Willman:2005b, Belokurov:2006, Belokurov:2007, Belokurov:2008, Belokurov:2009, Belokurov:2010, Zucker:2006, Walsh:2007}), the Dark Energy Survey (DES; \citealt{Bechtol:Drlica-Wagner:2015, Drlica-Wagner:Bechtol:2005, Kim:2015, Koposov:2015}), the Panoramic Survey Telescope and Rapid Response System Pan-STARRS1 (PS1; \citealt{Laevens:2015a, Laevens:2015b, Luque:2016}), and more recently with the Hyper Suprime-Cam Subaru Strategic Program (HSC-SSP; \citealt{Homma:2016, Homma:2018, Homma:2019}) and the DECam Local Volume Exploration Survey (DELVE; \citealt{Cerny:2021}). Modern simulations facilitate much higher mass resolution than before while utilizing complex baryonic physics prescriptions to describe important processes at the dwarf scale, including gas cooling, star formation, stellar feedback, cosmic rays, etc. \citep[e.g.,][among others]{Sawala:2016, Wheeler:2019, Vogelsberger:2020}.

These recent advances have given rise to a number of models for the connection between low-mass galaxies and dark matter halos \citep{GK2017:SMHM, Newton:2018, Jethwa:2018, Nadler:2019, Nadler:2020a, Kravtsov:2022, Manwadkar:2022}. Multiple studies have taken the approach of relating dark matter subhalos surrounding MW-like hosts predicted by simulations to the observed MW satellite galaxies. These studies assume a log-normal symmetric scatter, $\sigma$, in $M_\star$ at fixed $M_\mathrm{halo}$, about a median relation of the form $M_\star \propto M^\alpha_\mathrm{halo}$ at the low mass-end. 

Prior work that focused on the low mass-end SHMR has left a few puzzles. The first, nicely demonstrated in \citet{GK2017:SMHM}, is the degeneracy between the slope and the scatter about this relation. In particular, these authors show that a steep $M_\star-M^\alpha_\mathrm{halo}$ log-slope relation with a very high scatter ($\sim 2\,\mathrm{dex}$) and a shallow slope with zero scatter both results in similar predicted Local Group satellite stellar mass function. Secondly, by only utilizing the MW satellite population, we cannot ascertain whether the satellite SHMR is uniform for all hosts (at $z=0$) or unique for each host. In fact, studies often use the satellite population of the MW, which, in effect, only represents one ``realization'' of a true MW-like group, while using populations of dark matter subhalos from numerous realizations, each with its own unique evolutionary history. Answering this question is critical whenever ones use the satellite SHMR to constrain cosmological models. 

These questions can be better addressed with a large sample of well-characterized satellite galaxies around MW-like hosts beyond the MW. However, because of their low surface brightness, satellite galaxies around hosts external to the Local Group have proved challenging to survey in a comprehensive manner \citep{Danieli:2018}. Such surveys are now broadly possible thanks to new deep, wide-field imaging and spectroscopic surveys and new techniques for establishing reliable distances to satellite candidates and deriving their physical properties \citep{Merritt:2014, Karachentsev:2015b, Danieli:2017, Danieli:2020, Bennet:2017, Bennet:2019, Bennet:2020, Tanaka:2017, Smercina:2018, Cohen:2018, Garling:2021, Mutlu-Pakdil:2022}.

The Satellites Around Galactic Analogs Survey (SAGA; \citealt{Geha:2017, Mao:2021}) and the Exploration of Local VolumE Satellites (ELVES) Survey \citep{Carlsten:2019b, Carlsten:2020, Carlsten:2021a, Carlsten:2021b, Carlsten:2021c, Carlsten:2022}, are two surveys that systematically study the satellite populations around tens of massive galaxies beyond the Local Group. The two observational programs are complementary to each other as they target a different range in host luminosity, focus on groups at different distances, and differ in their sensitivity towards the faint end of the satellite luminosity function. Both surveys have the potential to offer the first constraints on the connection between satellite galaxies and dark matter subhalos, employing a large statistical sample of satellite galaxies beyond the MW. 

Here we present constraints on the connection between satellite dwarf galaxies and dark matter subhalos, using the new sample of dwarf satellites beyond the Local Group from the ELVES survey. The sample includes a total of $250$ confirmed and $71$ candidate dwarf satellites around 27 Milky Way (MW)-like hosts at $D<12\,\mathrm{Mpc}$. The galaxy-halo connection modeling requires two main ingredients, namely, the observed satellite galaxies and the simulated dark matter subhalos. In Section \ref{sec:observations} we describe this first ingredient, i.e., the observed sample of satellite galaxies from the ELVES survey, as well as the observational uncertainties that need to be considered. In Section \ref{sec:shmf} we provide details about the model used to predict the dark matter subhalos around MW-like hosts (\satgen) and how we construct the subhalo mass function. We describe the methodology and the statistical framework in Section \ref{sec:connection} and we present and discuss the results in Sections \ref{sec:results} and \ref{sec:discussion}, respectively.

\section{The Observed Satellite Galaxies Stellar Mass Function} \label{sec:observations}

The first key ingredient required for our analysis is the stellar mass function of satellite galaxies. In this section, we describe a sample of satellite dwarf galaxies in the Local Volume that was compiled and studied as part of the ELVES Survey (\S \ref{sec:elves}). We present the stacked ``Local Volume Satellite Stellar Mass Function'' (\S \ref{sec:SMF}), and we describe how we characterize the sample uncertainties and selection functions (\S \ref{sec:survey_charc}). 

\subsection{The ELVES Survey} \label{sec:elves}

The Exploration of Local VolumE Satellites (ELVES) Survey was designed to identify and characterize low-mass dwarf satellite galaxies surrounding massive host galaxies in the Local Volume \citep{Carlsten:2020,Carlsten:2021a,Carlsten:2021b,Carlsten:2021c, Carlsten:2022}. ELVES uses deep imaging data and customized techniques to comprehensively map the population of satellite galaxies surrounding 30 such hosts out to $D=12\,\mathrm{Mpc}$. A complete description of the survey is given in \citet{Carlsten:2022} and here we provide the details relevant to this work. 

The set of surveyed galaxy groups is unique in the sense that together they make a nearly volume-limited sample of MW-like hosts in the Local Volume. The groups are selected from two catalogs: the Updated Nearby Galaxy Catalog \citep{Karachentsev2013:groupcatalog} and the group catalog of \citet{KourkchiTully:2017}. Central galaxies are selected to have a distance of $D<12\,\mathrm{Mpc}$, a $K_s$-band magnitude of $M_{K_s} < -22.1\,\mathrm{mag}$ (with values taken from the \citet{KourkchiTully:2017} catalog), and a galactic latitude of $|b| > 17.4^\circ$ to avoid areas with high foreground extinction. Adopting a mass-to-light ratio of $M/L_{K_s}=0.6$ \citep{McGaughSchombert2014:masstolight}, the $K_s$-band magnitude threshold corresponds to a stellar mass of $M_\star\approx 10^{10}\,M_\odot$. These criteria result in 31 primary massive host galaxies, constructing a volume-limited sample of galaxy groups in a $R=12\,\mathrm{Mpc}$ sphere from us. 

30 of these 31 groups are surveyed for candidate satellites using archival Canada France Hawaii Telescope (CFHT)/MegaCam data (7 groups), data from the Dark Energy Camera Legacy Survey (DECaLS; \citealt{Dey:2019}) (17 groups), (non-DECaLS) DECam data (1 group) and by compiling lists of satellites from the literature (5 groups). Table 1 in \citet{Carlsten:2022} lists various properties of the surveyed groups. 

The candidate satellite identification is done by applying a semi-automated detection algorithm optimized for low surface brightness dwarf galaxies. The algorithm adopts notions from \citet{Bennet:2017} and \citet{Greco:2018}, as follows. The automated part of the detection pipeline uses a set of \texttt{SExtractor} \citep{Bertin:1996} runs with varying thresholds and aggressive masking of bright stars where needed. A visual inspection is performed to remove false detections such as light detected in the outskirts of massive galaxies and diffraction spikes from saturated stars that were not masked properly. 

Candidate satellites are confirmed as true satellites of their central host galaxy by assuring that their distance is roughly consistent with the group distance \citep[see][for the specific criteria used]{Carlsten:2022}. HST-based tip of the red giant branch (TRGB) distances and redshift measurements are used when available. However, distances to most candidates are measured through the surface brightness fluctuation (SBF) method \citep{Tonry:1988}. The SBF technique is calibrated for use in the case of dwarf galaxies in \citet{Carlsten:2019b} and \citet{Carlsten:2021c}, accounting for their stellar populations. The ground-based SBF measurements deliver $\sim 15 \%$ distance accuracy \citep{Carlsten:2019b} and in combination with the projected distance coverage of $300\,\mathrm{kpc}$, determine whether a satellite candidate is a true satellite or a background contaminant.
In total, for all 30 groups, out of 553 candidates discovered by ELVES, and an additional 87 satellites from the literature, 338 were confirmed as satellites (136 using the SBF technique and 202 through other methods), 196 were rejected as background contaminants, and 106 are remaining candidates (`unconfirmed') that did not have conclusive distance measurements. 

\citet{Carlsten:2022} provide an estimate for the probability that an `unconfirmed' candidate is a real satellite of its host central galaxy. The probability, $P_\mathrm{sat}$, is derived by comparing the candidates to other similar candidates that did have a conclusive distance measurement. In particular, we calculate the likelihood that a candidate is a real satellite as a function of its $V$-band absolute magnitude ($M_V$) and central surface brightness ($\mu_{0,V}$). The likelihood is estimated from the fraction of confirmed satellites out of the closest 20 confirmed or rejected candidates for each point on the $M_V - \mu_{0,V}$ plane (see Section 5.6 and Figure 5 in \citealt{Carlsten:2022}). As described in Section \ref{sec:SMF}, we use the estimated $P_\mathrm{sat}$ for each candidate when deriving the satellite stellar mass functions for the ELVES groups.

The optical photometry was obtained by fitting a two-dimensional \sersic model to the DECaLS/DECam or CFHT data using the $g$, and $r$ or $i$ bands as described in \citet{Carlsten:2021b} and \citealt{Carlsten:2022}. The $g$-band luminosity and either the $g-r$ or $g-i$ colors are used to estimate the stellar masses, using the color-dependent mass-to-light ratio relations from \citet{Into:2013}. In some cases, where a single \sersic model cannot provide a good fit to the galaxy, the stellar mass is calculated from the Two Micron All Sky Survey (2MASS; \citealt{Skrutskie:2006}) $M_{K_{s}}$ values \citep{KourkchiTully:2017}.

\begin{figure*}[t!]
 \centering
 \includegraphics[width=1.\textwidth]{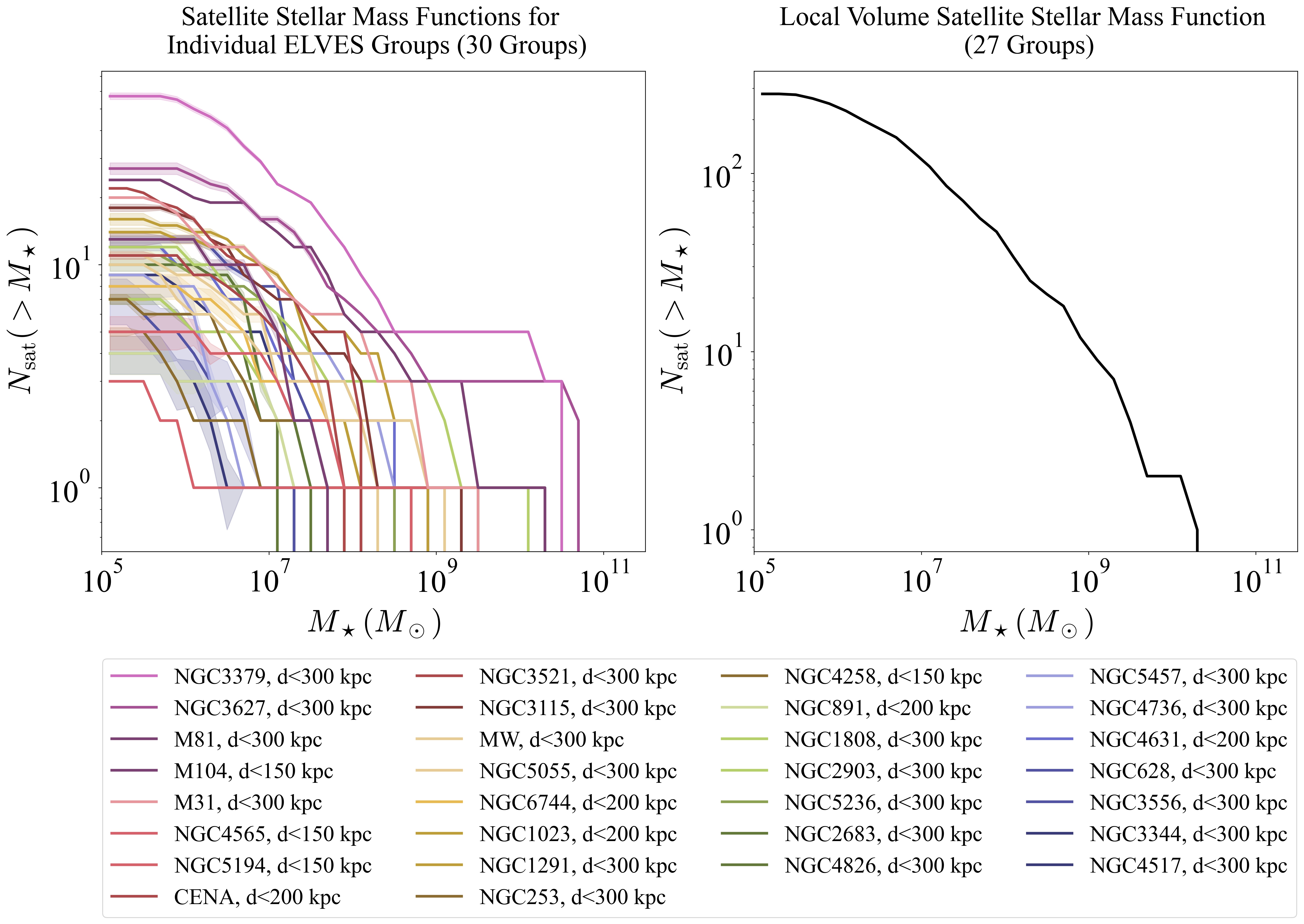}
  \caption{\textit{Left:} The satellite stellar mass function (SSMF) for individual groups in the ELVES Survey where each curve represents the SSMF for one Local Volume group surveyed in ELVES. We forward-model completeness in Section \ref{sec:modelsmf} but do not correct it here. The shaded regions associated with each curve show the $1\sigma$ uncertainty of the number of satellites, taking into account the `unconfirmed/candidate' satellites (see text). The groups are listed in the legend sorted by their $K$-band magnitude (with NGC3379 having the brightness group $K$-band magnitude of $M_{K,\mathrm{NGC3379}} = -25.37\,\mathrm{mag}$ and NGC4517 having the faintest group $K$-band magnitude of $M_{K,\mathrm{NGC4517}} = -22.16\,\mathrm{mag}$. The legend also lists the radial coverage (in projected kpc from the group central galaxy) of the survey imaging used to find candidate satellites. \textit{Right:} The combined SSMF for the 27 groups used in the analysis (excluding the NGC3379, NGC3627, and M81 groups). In total, the black curve shows $279 \pm 4$ satellite galaxies and are used in the analysis.}
  \label{fig:elves_smf}
\end{figure*}

\subsection{The ``Local Volume Satellites Stellar Mass Function''} \label{sec:SMF}

The ELVES catalog of confirmed and unconfirmed satellite galaxies is used to construct the satellite stellar mass function (SSMF) for the groups studied in the ELVES survey. For each individual host, we calculate its SSMF by considering its confirmed satellites and also its unconfirmed/candidate satellite with their associated probabilities (Section \ref{sec:elves}). We calculate the stellar mass function 10,000 times for each group, where in each realization the confirmed satellites are always included (with probability 1) and the choice of whether to include an unconfirmed/candidate satellite is made based on its $P_\mathrm{sat}$. For example, a candidate satellite with $P_\mathrm{sat}=0.9$ will be included in $90\%$ of the realizations.

The left panel of Figure \ref{fig:elves_smf} shows the median SSMFs for the 30 ELVES groups (solid curves) down to $M_\star=10^5\,M_\odot$. The shaded regions bracketing each curve show their standard deviation using the 10,000 realizations. As can be seen, the unconfirmed/candidate satellites addition is small for most groups, ranging between 0 and 2 satellites per group. The legend below shows the central galaxy names for each group, sorted by their $K_s$-band magnitude, and the group radial coverage out to which candidate satellites were searched. The three brightest central galaxies are NGC3379, NGC3627, and M81, which are also the first three in the $K_s$-band magnitude-ranked figure legend. They host the largest number of confirmed satellites -- 57, 27, and 24, respectively. Interestingly, these three groups have the largest number of massive satellites and as mentioned in \citet{Carlsten:2022}, their group luminosity is significantly larger than that of the primary host. As explained below, we exclude these three `small groups' from the analysis as their higher mass, and likely different formation history from bonafide MW-analogs may bias the results. 

The goal of this paper is to use the collective information about the satellite galaxies in the Local Volume to put constraints on the satellite SHMR. Therefore, we combine the individual SSMFs from the 27 Local Volume groups into one, ``Local Volume Satellite Stellar Mass Function'', as shown in the right panel of Figure \ref{fig:elves_smf}. Similarly to the individual group SSMFs, we calculate the median and standard deviation Local Volume SSMF from 10,000 realizations where the unconfirmed/candidate satellites are included statistically, based on their association probability. In total, the black curve shows $279 \pm 4$ satellite galaxies used in the analysis.

\begin{figure*}[t]
 \centering
 \includegraphics[width=1.\textwidth]{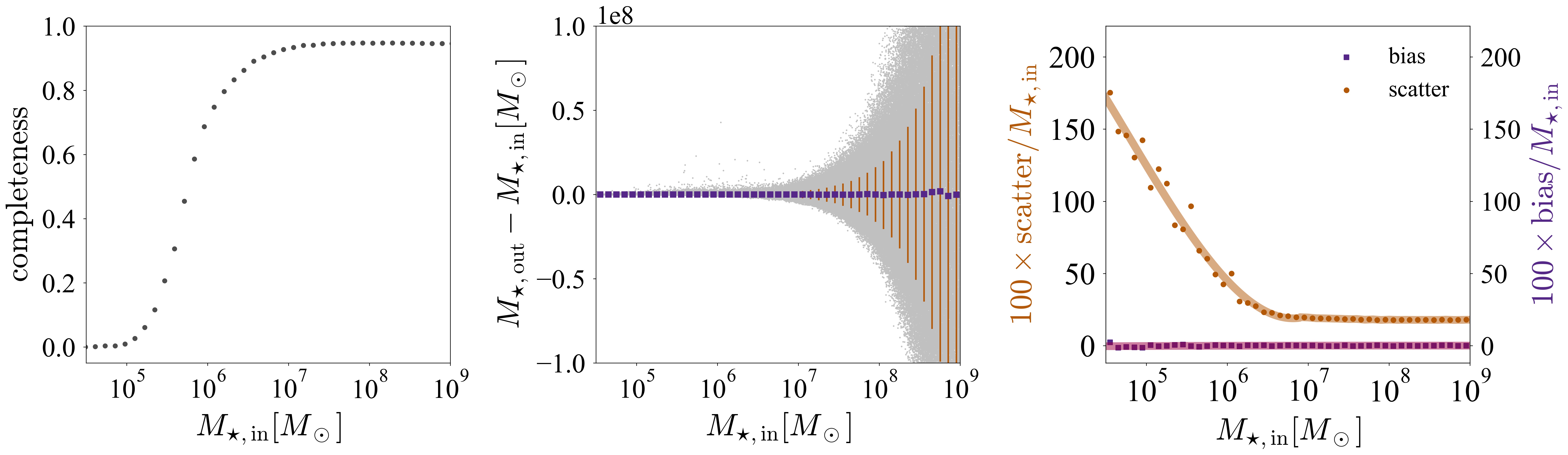}
  \caption{The completeness, bias, and scatter as a function of stellar mass, of the ELVES satellite galaxies. The left panel shows the completeness quantified by injecting artificial galaxies into the real survey images. The two rightmost panels show the systematic offset (bias; shown in purple) and the scatter (orange) between the intrinsic (true) and recovered stellar masses. The relations are fitted with a third-order polynomial and used to forward model the satellite stellar mass function in Section \ref{sec:modelsmf}.}
  \label{fig:survey_charac}
\end{figure*}

\subsection{Sample Characterization -- Completeness, Stellar Mass Bias and Scatter} \label{sec:survey_charc}

Understanding how the survey and sample limitations impact the observables used in the analysis is crucial. Moreover, such understanding is particularly important when relying on this sample of dwarf satellites for constraining theoretical models, as we aim to do in this work. The three main sources of uncertainty are

\begin{enumerate}
    \item $\mathrm{completeness}(M_\star)$: the sample incompleteness as a function of the stellar mass.
    \item $\mathrm{bias}(M_\star)$: the systematic error (``bias'') in the stellar mass.
    \item $\mathrm{scatter}(M_\star)$: the statistical photometric scatter in the stellar mass.
\end{enumerate}

\noindent
where the latter two are driven by photometric errors in the measured luminosities and colors used to estimate the stellar mass. We expect that the bright dwarf galaxies will have a very small scatter and that the photometric error will increase towards fainter galaxies.

To quantify the sample completeness, we use the results of the injection simulations performed in \citet{Carlsten:2022}. Briefly, artificial galaxies are injected into the survey images, spanning a range of $r$/$i$-band absolute magnitudes and $g-r$/$g-i$ colors (depending on what filters were used for the imaging). Where possible, artificial galaxies are injected at the chip level before sky subtraction and image co-addition. This assures that possible inadvertent subtraction of low surface brightness galaxies as part of the background subtraction step is taken into account. After injection, the images are run through the same detection pipeline, and the fraction of recovered galaxies is calculated as a function of input magnitude and surface brightness (or size, as all injected galaxies were generated with an $n=1$ S\'{e}rsic profile). Artificial galaxies that are injected onto star-masked pixels are taken into account as well; the area lost to masked pixels is around $5\%$. A thorough description of the completeness analysis and results are described in Section 4.2, Appendix C, and Figure 4 in \citet{Carlsten:2022}.

We use the results presented in \citet{Carlsten:2022} to derive a one-dimensional function for $\mathrm{completeness}(M_\star)$. In particular, Figure 4 in \citet{Carlsten:2022} shows the detection efficiency as a function of $M_g$ and $\mu_{0,g}$, averaged across 25 hosts\footnote{The recovery fractions are similar across the various hosts and imaging data and therefore taking the average is sensible.}. To convert to $\mathrm{completeness}(M_\star)$, we do the following. First, we sample galaxies across a range in stellar mass and assign them sizes using the mass-size relation that ELVES satellites follow \citep{Carlsten:2021b} while accounting for the intrinsic scatter in size at fixed $M_\star$ ($\sigma \sim 0.2\,\mathrm{dex}$; \citealt{Carlsten:2021b}). We then assign a color to each galaxy by sampling from the overall color distribution of the ELVES satellites. Then, we get $M_g$ and $\mu_{0,g}$ for each simulated galaxy assuming the color-$M/L$ relation of \citet{Into:2013} and, finally, we read the probability that this galaxy will be detected from the left panel of Figure 4 in \citet{Carlsten:2022}. We note that we remove galaxies fainter than $M_V=-9\,\mathrm{mag}$ as is done with the observed satellites catalog. The completeness as a function of the stellar mass is shown in the left panel of Figure \ref{fig:survey_charac}. Above $M_\star \sim 2 \cdot 10^7\,M_\odot$, the completeness is maximal at 95\%. As we noted earlier, $\sim 5\%$ of the area is lost due to masked pixels. The completeness is $50\%$ for $M_\star \sim 5.7\cdot 10^5\,M_\odot$ and it drops steeply below this mass, likely due to the sharp cut at $M_V=-9\,\mathrm{mag}$.

Next, we quantify the systematic errors in estimating the stellar mass (``bias'') and the photometric scatter, as follows. We simulate 1,000,000 galaxies with an absolute magnitude, $M_g$, drawn uniformly between $-20$ and $-8\,\mathrm{mag}$. Each galaxy also gets assigned a distance drawn from the distribution of distances to the ELVES satellites, a size, $r_\mathrm{eff}$ drawn from the ELVES $M_g - r_\mathrm{eff}$ relation, and color drawn from the color distribution of the ELVES dwarfs. We then calculate the intrinsic (``true'') apparent magnitude of each simulated dwarf assuming its assigned distance and absolute magnitude. We also calculate the true stellar mass of each simulated dwarf, $M_{\star,\mathrm{in}}$ from its absolute magnitude and color.
Next, to estimate the recovered stellar mass, $M_{\star,\mathrm{out}}$, we use the relations derived in \citet{Carlsten:2021b}, shown in Figure 17 in that work, that give estimated errors for $m_g$ and the color, $\Delta m_g|_{m_g}$ and $\Delta \mathrm{color}|_{m_g}$, respectively. For each simulated dwarf, we perturb its apparent magnitude by $\Delta m_g|_{m_g}$ and color by $\Delta \mathrm{color}|_{m_g}$ and finally, calculate $M_{\star, \mathrm{out}}$, the recovered stellar mass. 

The middle and right panels of Figure \ref{fig:survey_charac} show the results. The stellar mass bias (purple) is small across all stellar masses, ranging in most cases between $0$ and $\sim 2-3\%$. On the other hand, the scatter (orange) is much more significant, with $\sim 20\%$ down to $M_\star \sim 5 \cdot 10^6\,M_\odot$, and keeps increasing rapidly below this mass. We quantify the bias as a function of stellar mass by fitting a third-order polynomial, shown by the solid purple line. To estimate the scatter as a function of stellar mass, we interpolate the simulation results for masses $M_{\star, \mathrm{in}}>10^7\,M_\odot$, and fit a third-order polynomial to masses $ M_{\star, \mathrm{in}}<10^7\,M_\odot$. The resulting model for the scatter is shown by the solid orange line. In Section \ref{sec:modelsmf}, we use these derived relations to forward model the observed satellite stellar mass function.

\section{The Subhalo Mass Function}\label{sec:shmf}

The second key ingredient is the predicted population of dark matter subhalos orbiting the Local Volume host galaxies. These are needed in order to model the satellite-subhalo connection. In this section, we describe how we model the population of subhalos for each ELVES group using the \satgen code (\S \ref{sec:satgen}) and the construction of a joint ``Local Volume Subhalo Mass Function'' (\S \ref{sec:lvshmf}).

\subsection{The SatGen Model} \label{sec:satgen}

\satgen is a publicly available semi-analytical dark matter halo and satellite galaxy generator\footnote{\url{https://github.com/shergreen/SatGen}}. The model and examples are described in detail in \citet{Jiang2021:SatGen} and \citet{Green2021:SatGenApp}, and here we provide a brief description of the central concepts and our application of the subhalo generator. \satgen offers multiple improvements compared to past models. First, it includes the effects of baryonic physics on both the host halo and its subhalo population. Such effects have been shown to play an important role on the subhalos' abundance, their spatial distribution within their host halo, as well as the structure of individual subhalos (e.g. \citealt{GK:2019}). Second, the prescriptions used in determining the subhalos' structural evolution are adopted either from high-resolution idealized simulations or using physically-motivated models and thus do not suffer from the effects of numerical, artificial disruption of subhalos, as identified in cosmological simulations \citep{vdb:2018a, vdb:2018b}. Finally, \satgen is computationally cheap and can be easily used to simulate many hundreds of host halos while still maintaining adequate resolution for studying low mass subhalos galaxies ($M_\mathrm{sub} \gtrsim 10^8\,M_\odot$). Given a host halo of mass, $M_\mathrm{halo}$, and redshift, $z$, \satgen generates halo merger trees, using an algorithm based on the extended Press-Schechter (EPS) formalism \citep{Lacey1993:EPS}, taking into account a corrective factor for the low-redshift merger rate as presented in \citet{Parkinson2008:Tree}. The host dark matter halo profile is initialized and the halo response model is set to match the response model from state-of-the-art cosmological simulations, chosen by the user. As described in \citet{Jiang2021:SatGen}, the halo response model can be chosen from two different models which are representative of simulations of bursty star formation and strong supernovae feedback, such as NIHAO \citep{Wang:2015} and FIRE \citep{Hopkins:2014}, and of simulations of non-bursty star formation and weaker feedback, such as APOSTLE \citep{Sawala:2016} and Auriga \citep{Grand:2017}.

Then, the initial conditions for the satellite galaxies are set, including the properties of the host system when the satellites enter the virial sphere, the orbit of the incoming satellites, and the dark matter, stellar and gaseous properties of the incoming satellites. Finally, \satgen evolves the satellites by integrating their orbits, taking into account the effects of dynamical friction, tidal stripping, and ram pressure stripping, using analytical prescriptions. 

For our fiducial run, we set up the \satgen model as follows. We choose the host potential to be a dark matter halo following the Dekel+ profile \citep{Dekel2017:DMProfile} with an embedded galactic disk that makes up $10\%$ of the mass of the halo. We assume bursty star formation and strong supernovae feedback, emulating the halo response as done in simulations such as NIHAO \citep{Wang2015:NIHAO} and FIRE \citep{Hopkins2018:FIRE}. As mentioned above, the other option allowed by \satgen for a halo response is non-bursty star formation and weaker feedback, such as done in the APOSTLE \citep{Sawala:2015} and Auriga \citep{Grand:2017} simulations. \citet{Jiang2021:SatGen} showed that the effect of choosing one halo response model or the other is minor, where the NIHAO-like feedback yields $\sim 5\%$ fewer satellites than the APOSTLE-like model. Therefore, we do not explore the non-bursty option and leave a more thorough exploration of this effect to future work.

Using this setup, we create a library of \satgen runs for $z=0$ host halos with virial masses ranging from $M_{\mathrm{vir}}=10^{10.5}\,M_\odot$ to $M_{\mathrm{vir}}=10^{13.3}\,M_\odot$ in $\log M_{\mathrm{vir}}=0.01$ jumps (with a total of 280 \satgen runs). For each input host halo virial mass, we generate a merger tree recording progenitor halos down to $10^{8}\,\mathrm{M}_\odot$, initialize the host and satellites, and evolve the satellites to z=0. While \satgen provides a full description of the evolving satellite galaxies (e.g. their stellar mass, size, etc.), we only use the details related to the dark matter subhalos (e.g. the peak mass) and do not use any information related to the satellite galaxies themselves.

\subsection{Constructing a ``Local Volume Subhalo Mass Function''} \label{sec:lvshmf}

We construct one combined ``Local Volume Subhalo Mass Function'' for the 27 LV hosts modeled with \satgen and included in our analysis, in the following way. For each of the 27 hosts, we calculate its stellar mass by using its group $K_s$-band luminosity \citep{KourkchiTully:2017} and adopting a mass-to-light ratio of $M/L_{K_s}=0.6$ \citep{McGaughSchombert2014:masstolight}. In the majority of cases, the group $K_s$ luminosity is identical to the host luminosity, however, for groups with more than one massive galaxy, it is significantly higher. Therefore, adopting the group $K_s$ luminosity rather than the central galaxy $K_s$ luminosity provides a more accurate estimate of the total stellar mass enclosed in the host halo. To get the halo mass for individual ELVES hosts, we use then use the \citet{RP:2017} SHMR at $z=0$, assuming the stellar mass estimated for each group and a constant scatter of $0.15\,\mathrm{dex}$ \citep{RP:2015}.

\begin{figure*}[t!]
 \centering
 \includegraphics[width=1.\textwidth]{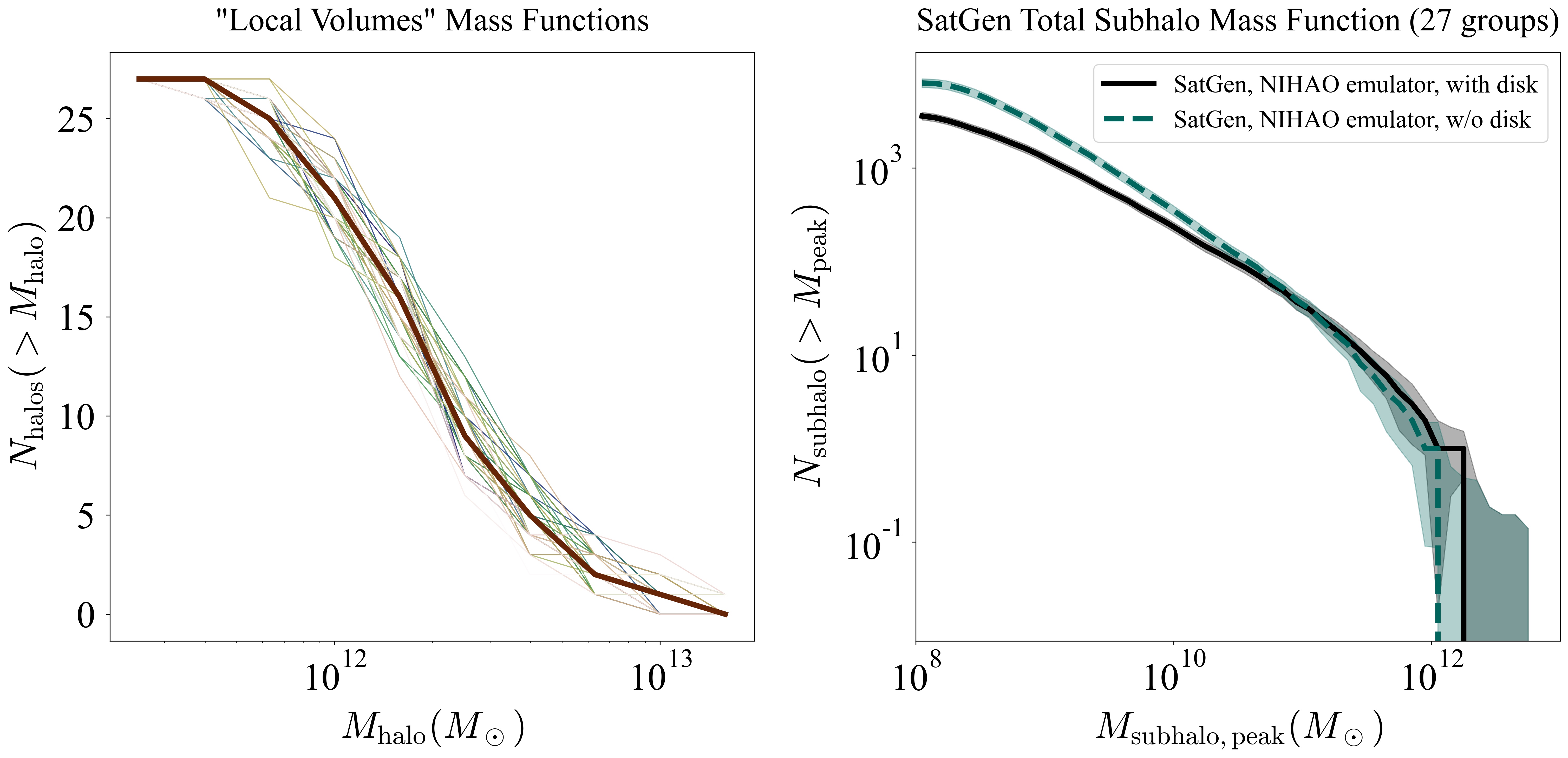}
  \caption{Left panel: the halo mass functions for different “Local Volumes”, each with 27 halos. The halo masses are sampled from a normal distribution around the mean SHMR for massive galaxies from \citet{RP:2017} with a constant scatter of 0.15. Right panel: the ``total'' subhalo mass function using \satgen for all 27 hosts together. The black curve is the median and the shaded black is one standard deviation.
}
  \label{fig:shmf}
\end{figure*}

One Local Volume realization is constructed by drawing the halo masses of its 27 members assuming the stellar mass of each host galaxy. Each such volume realization is composed of slightly different halo masses for its members, accounting for the scatter in the SHMR for massive galaxies. In the left panel of Figure \ref{fig:shmf}, we show the mass function of 50 such volumes. We then use the precalculated \satgen library to compile the census of subhalos in each volume by including the subhalo catalog from the \satgen run that is the closest in host halo mass. In the right panel of Figure \ref{fig:shmf}, we show the ``total'' subhalo mass function constructed using the \satgen library for all 27 hosts. The solid black line shows the median subhalo mass function of the 50 Local Volume realizations using our fiducial \satgen model (see Section \ref{sec:satgen}) and the shaded regions show the volume-to-volume variance (one standard deviation).

\section{Galaxy-Halo Connection}\label{sec:connection}

\subsection{Methodology Overview} \label{sec:method}

\begin{figure*}[t!]
 \centering
 \includegraphics[width=1.\textwidth]{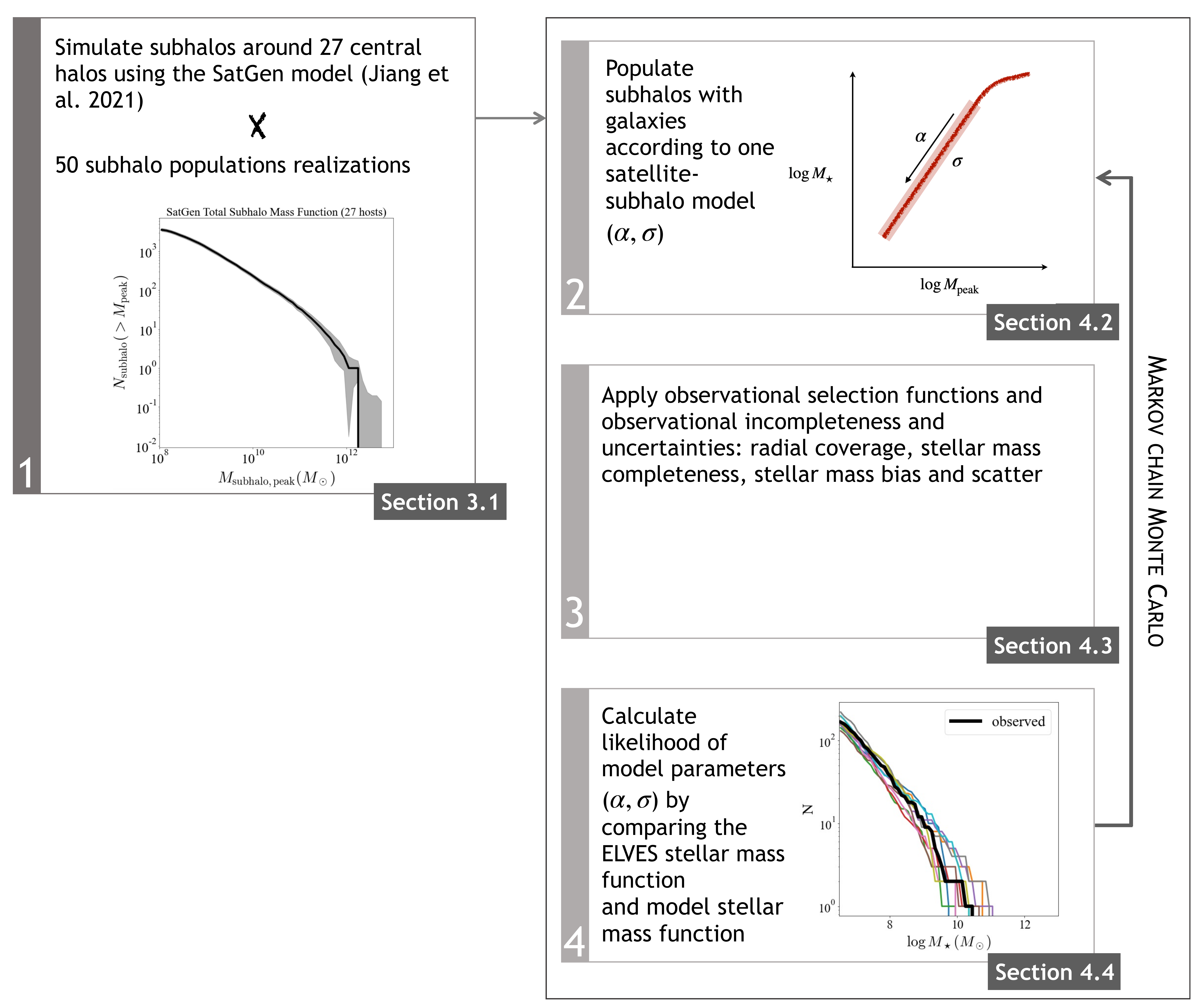}
  \caption{The methodology used in this work to constrain the satellite stellar-to-halo mass relation (\S \ref{sec:method}). We utilized the \satgen model to simulate 50 realizations of the ``Local Volume'' subhalos around 27 host galaxies (Step 1, \S \ref{sec:satgen}). Stellar masses are assigned to dark matter subhalos using a four-parameter SHMR relation (Step 2, \S \ref{sec:model}), and the stellar mass function is forward-modeled to incorporate the observational selection functions and observational incompleteness and uncertainties (Step 3, \S \ref{sec:modelsmf}). In step 4 (\S \ref{sec:mcmc}), the modeled satellite stellar mass function is compared to the observed ``Local Volume'' satellite mass function, and the likelihood for the chosen model parameters is calculated. Steps 2-4 are repeated until the posterior distribution is computed, spanning the pre-determined priors.}
  \label{fig:diagram}
\end{figure*}

Having the observed Local Volume satellite stellar mass function and the simulated population of dark matter subhalos, we now turn to describe the methodology we use to model the subhalo-satellite connection. Our method is visually illustrated in Figure \ref{fig:diagram}. The steps are as follows:

\begin{enumerate}
    \item 50 realizations of the subhalos around the 27 ELVES host galaxies are simulated using the \satgen code, as described in Section \ref{sec:satgen}. This step is performed only once.
    \item Galaxies are assigned to subhalos according to the satellite-subhalo connection model, characterized by six parameters -- $\log \epsilon$, $M_1$, $\alpha$, $\sigma$, $\delta$, and $\gamma$, where only two parameters are allowed to change, $\alpha$ and $\sigma$, as described in \ref{sec:model}.
    \item The observational selection functions are applied and observational uncertainties (the sample incompleteness, and the bias and scatter in the derived stellar mass) are forward modeled (\S \ref{sec:modelsmf}).
    \item The forward-modeled stellar mass function is compared to the observed one (\S \ref{sec:SMF}) and the likelihood for the satellite-subhalo connection model parameters chosen in step 2 is calculated.
    \item Steps 2-4 are repeated until the posterior distribution is well-sampled. The parameter space is explored using the Markov Chain Monte Carlo (MCMC) method. 
    
\end{enumerate}

A similar approach was taken in e.g. \citet{Jethwa:2018} and in \citet{Nadler:2019, Nadler:2020a}. The substantial differences are the input catalog of observed satellite galaxies and the way the dark matter subhalos are modeled.

In the following subsections, we present the satellite-subhalo connection model parameterization (\S \ref{sec:model}), our procedure to forward-model the ``observed'' stellar mass function, taking into account the observational selection functions (\S \ref{sec:modelsmf}), and the statistical framework we adopt to estimate the posterior distribution on the satellite–subhalo connection model parameters (\S \ref{sec:mcmc}).

\subsection{Satellite-Subhalo Connection Model}\label{sec:model}

We parameterize the relation between the stellar mass of the satellites, $M_{\star}$, and their host subhalo peak masses, $M_\mathrm{peak}$, using the functional form presented in \citet{Behroozi:2019}\footnote{$M_\mathrm{peak}$ is adopted for the halo masses, as is regularly used in halo occupation studies.}. This work, utilizing the U\textsc{niverse}M\textsc{achine} code, provides fitting formulae for median SHMR at 22 redshifts with best-fitting parameters for different selection cuts. The exact parametrization of the $M_\star - M_\mathrm{peak}$ relation we adopt is:

\begin{equation} \label{eq:model_median_mstar}
\begin{split}
\log_{10} M_{\star} = \epsilon + \log_{10}(M_1) -\log_{10}(10^{-\alpha x} + 10^{-\beta x}) \\
+ \gamma \exp\left[{-0.5\left(\frac{x}{\delta}\right)^2}\right]
\end{split}
\end{equation}

with 

\begin{equation}
x = \log_{10}\left(\frac{M_\mathrm{peak}}{M_1}\right)
\end{equation}

This functional form for the SHMR is a double power law \citep{Behroozi:2010, Behroozi2013, Moster2013} plus a Gaussian, with a characteristic peak halo mass ($M_1$), a characteristic stellar mass to peak halo mass ratio, $\epsilon=\frac{M_\star|_{M_\mathrm{
peak}=M_1}}{M_1}$, a faint-end slope of $\alpha$, a massive-end slope of $\beta$, the height of the Gaussian, $\gamma$, and the Gaussian width, $\delta$.

We model the scatter in $M_\star$ by a symmetric, log-normal distribution at fixed $M_\mathrm{peak}$. For our fiducial model, we adopt a constant scatter as a function of $M_\mathrm{peak}$. We also examine and discuss the results of a ``growing scatter'' model in Section \ref{sec:growing_scatter}.

We fix all model parameters to the best-fit values from \citet{Behroozi:2019}, except for the faint-end slope, $\alpha$, and the scatter, $\sigma$. In particular, we adopt the \citet{Behroozi:2019} best-fit parameters for the median SHMR at $z=0$ for satellites galaxies (without a distinction between star-forming and quenched galaxies), taking: \footnote{Values are taken from Table J1 in \citet{Behroozi:2019}}: $\epsilon =-1.432$, $M_1 = 11.889$, $\beta = 0.464$, $\log_{10}(\gamma) = -0.812$, and $\delta = 0.319$.

In the following sections, we describe the procedure of estimating the posterior probability of the two free model parameters: $\alpha$ and $\sigma$.

\subsection{Forward Modeling of the Observed Stellar Mass Function}\label{sec:modelsmf}

\begin{figure*}[t]
 \centering
 \includegraphics[width=1.\textwidth]{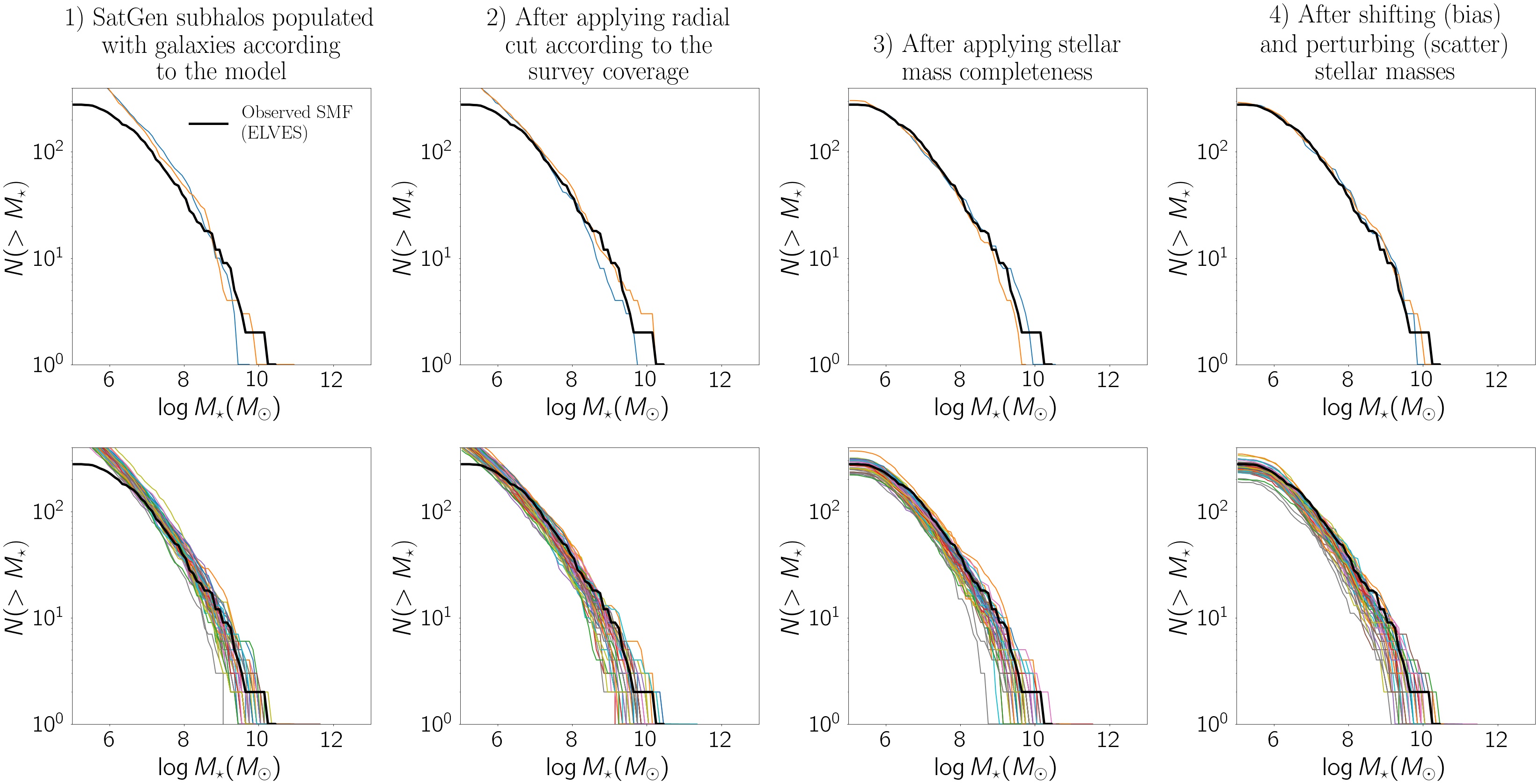}
  \caption{Forward modeling the satellite stellar mass function used in inferring the SHMR model parameters. The black curve in all panels shows the observed ELVES satellite stellar mass function, compared to the colored curves showing the modeled stellar mass functions in each one of the steps. The upper row shows only two such modeled stellar mass functions for clarity whereas the lower row shows all 50 simulated ``Local Volumes''.}
  \label{fig:model_smf}
\end{figure*}

With the sample completeness and errors parameterized (\S \ref{sec:survey_charc}) and the SHMR model functional form in place (\S \ref{sec:model}), we forward model the satellite stellar mass function, as follows. 
First, we assign satellite galaxies with a stellar mass $M_\star$ to modeled \texttt{SatGen} subhalos with $M_\mathrm{peak}$, according to a satellite-subhalo connection model described in Section \ref{sec:model}. 
Next, a radial cut is performed to only include simulated satellites in a way that mimics the radial coverage of the ELVES survey. As mentioned earlier, not all ELVES groups are surveyed out to $300\,\mathrm{kpc}$ due to limited coverage in some of the data sets used. We, therefore, rank the ELVES hosts by their $K_s$ group luminosity and the \satgen halos by their $M_{200}$ and match their radial coverage. For example, if an ELVES group was only searched for satellites out to $200\,\mathrm{kpc}$ (like in the case of NGC4631), only the subhalos associated with its matched \satgen halo with $r_\mathrm{proj}<200\,\mathrm{kpc}$, viewed from random viewing sight-lines, will be kept and contribute to the modeled stellar mass function. Then, the sample incompleteness derived in Section \ref{sec:survey_charc} (including both the incompleteness due to the surface brightness limit and the cutting out galaxies fainter than $M_V=-9~\mathrm{mag}$) is applied to the intrinsic stellar mass of the simulated satellites. Finally, the observational errors characterized in Section \ref{sec:survey_charc} are applied to the intrinsic stellar masses. We shift the stellar mass by the parameterized bias and perturb it by a random number drawn from a Gaussian distribution with a width that is given by the parameterized scatter.

The forward-modeled satellite stellar mass function through each of these four modeling steps is shown in Figure \ref{fig:model_smf}. The black curve shows the observed stellar mass function from ELVES and the colored curves show stellar mass function models from various \satgen realizations, adopting one set of model parameters (not the ``best-fit'' parameters). The upper row shows just two such realizations and the lower row shows all 50 used in the analysis.

\subsection{Statistical Framework}\label{sec:mcmc}

Constraining the satellite-subhalo connection model parameters hinges on comparing the discrete observed and modeled satellite samples. We adopt the statistical framework described in \cite{Nadler:2019,Nadler:2020a}, assuming the observed satellites and the simulated satellites populate the stellar mass bins according to a Poisson point process. 

Let $n_{\mathrm{obs},i}$ be the observed number of satellites in stellar mass bin $i$, and $n_{\mathrm{model},i,j}$, be the number of modeled satellites in stellar mass bin $i$, in a Local Volume realization $j$. The likelihood of observing the ELVES satellite stellar mass function, $\mathrm{E}_{\mathrm{ELVES}}$, given a particular satellite-subhalo connection model parameters, $\theta$ is 

\begin{equation} \label{eq:likelihood}
    P(\mathrm{E}_{\mathrm{ELVES}} | \theta) = \prod_{i=1}^{N_{\mathrm{bins}}} P(n_{\mathrm{obs},i} | n_{\mathrm{model},i,j})
\end{equation}

Marginalizing over the Poisson rate parameter (the expected number of satellites in each bin), $\lambda_i$ (see \citealt{Nadler:2019, Nadler:2020a}), we have

\begin{equation}
\begin{split}
    &P(n_{\mathrm{obs},i} | n_{\mathrm{model},i,1}, ..., n_{\mathrm{model},i,N}) = \\
    &\left( \frac{N+1}{N} \right) ^{-(n_{\mathrm{model},i,1} + \cdot\cdot\cdot + n_{\mathrm{model},i,N} + 1)} \nonumber \\
    &\times (N+1)^{-n_{\mathrm{obs},i}} \frac{(n_{\mathrm{obs},i}+n_{\mathrm{model},i,1} + \cdot\cdot\cdot + n_{\mathrm{model},i,N} )!}{n_{\mathrm{obs},i}! (n_{\mathrm{model},i,1} + \cdot\cdot\cdot + n_{\mathrm{model},i,N})!} 
\end{split}
\end{equation}

where $N=50$ is the number of ``Local Volume'' realizations and we assume that $n_{\mathrm{obs},i}$ and $n_{\mathrm{model},i,j}$ are drawn from the same Poisson distribution with the same rate parameter. We use Bayes's theorem to compute the posteriors for the satellite-subhalo model:

\begin{equation}
    P(\theta | \mathrm{E}_{\mathrm{ELVES}}) = \frac{P(\mathrm{E}_{\mathrm{ELVES}} | \theta)  P(\theta)}{P(\mathrm{E}_{\mathrm{ELVES}})}
\end{equation}

\noindent
where $P(\mathrm{E}_{\mathrm{ELVES}} | \theta)$ is the likelihood function defined in equation \ref{eq:likelihood}, $P(\theta)$ are the prior distributions, and $P(\mathrm{E}_{\mathrm{ELVES}})$ is the Bayesian evidence corresponding to the observed ELVES satellite sample.

We use the Markov Chain Monte Carlo (MCMC) sampler \texttt{emcee} \citep{FM2013:emcee} to sample the posterior for the fit to the ELVES satellite mass function. For our fiducial parameter inference, we use the simulated satellite populations from 50 ``Local Volume'' realizations as described in \ref{sec:lvshmf}. We assume the following priors, informed by previous work \citep{Jethwa:2018, Nadler:2019, Nadler:2020a, Munshi2021:SMHM}: $\alpha \sim \mathrm{unif}(0,3)$ and $\sigma \sim \mathrm{unif}(0,2)$.
We run $5 \times 10^4$ iterations using 20 walkers, discarding the first 5000 burn-in steps. The MCMC run yielded an acceptance rate of $\mathrm{f}_\mathrm{accept} = 0.35$ and increasing the number of walkers and iterations did not change the results.

\section{Modeling Results}\label{sec:results}

\subsection{The inferred fiducial model parameters} \label{sec:results_fiducial}

\begin{figure*}
\centering     
\subfigure{\label{fig:a}\includegraphics[width=85mm]{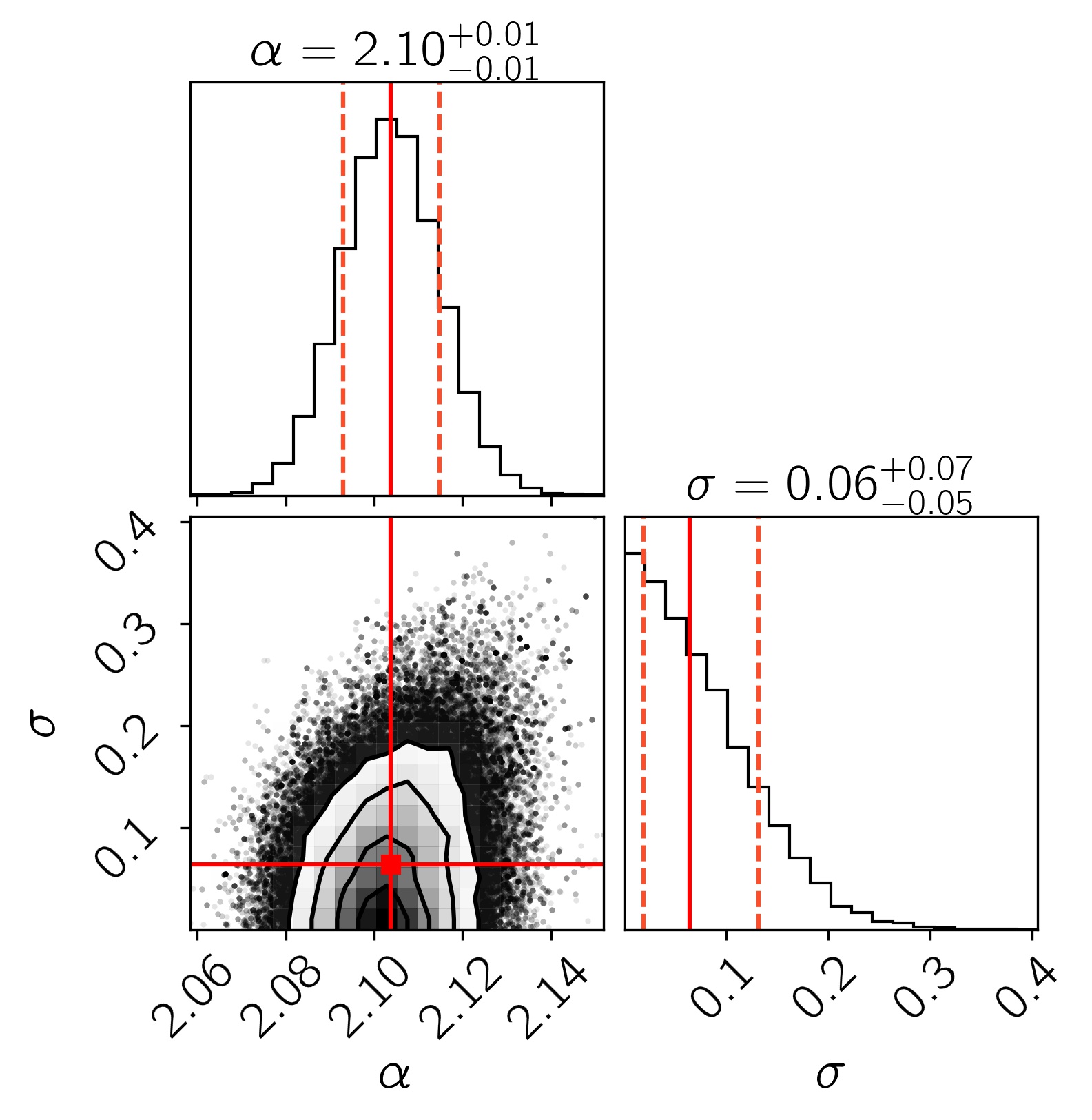}}
\subfigure{\label{fig:b}\includegraphics[width=85mm]{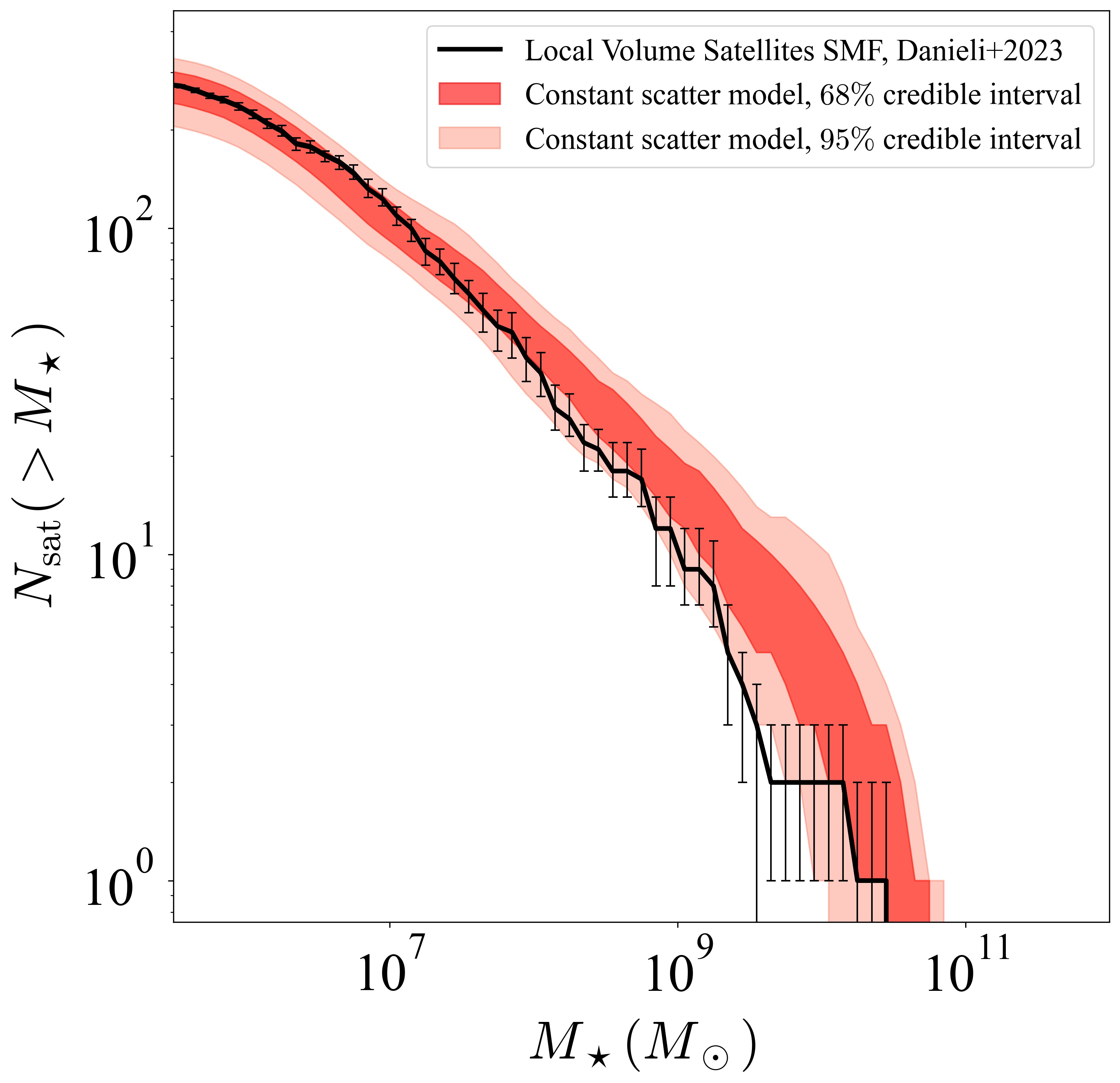}}
\caption{Results for the fiducial satellite-subhalo connection model, assuming a constant scatter. \textit{Left panel:} Posterior distribution of the parameters in the constant scatter SHMR model, derived using 321 satellites and satellite candidates around 27 host galaxies in the Local Volume. Solid red lines indicate the median (50\%) and dashed red lines mark the 68\% credible interval. See Table \ref{table:results} for parameter values. \textit{Right panel:} The 68\% (red) and 95\% (light red) credible intervals for the satellite stellar mass function predicted by the constant scatter model, compared to the satellite stellar mass function observed with ELVES (black).}
\label{fig:constant_scatter}
\end{figure*}

The posterior distributions of the two model parameters, $\alpha$ and $\sigma$ are shown in the left panel of Figure \ref{fig:constant_scatter}. The vertical, solid red line denotes the median value and the vertical dashed lines indicate the $68\%$ credible intervals. The $68\%$ and $95\%$ credible intervals are summarized in Table \ref{table:results}.

The inferred faint-end slope of the $M_\mathrm{peak}-M_\star$ relation below the characteristic peak halo mass $M_1$ is $\alpha = 2.10 \pm 0.01$. The constant scatter is inferred to be $\sigma=0.06^{+0.07}_{-0.05}$. We note that this value represents the posterior median while the posterior distribution gives only an upper limit on the scatter as can be seen in the left panel of Figure \ref{fig:constant_scatter}. This indicates that the observed SSMF does not require significant scatter in the $M_\star - M_\mathrm{peak}$ relation in the dwarf regime. The constant scatter, $\sigma$ is constrained to be lower than $0.30$ at three sigmas ($99.85$ percentile).

In the right panel of Figure \ref{fig:constant_scatter}, we compare the observed SSMF with the $68\%$ (red) and $95\%$ (pink) credible intervals for the stellar mass function, drawn from the posterior distributions. The predictions for the satellite galaxies around the 27 ELVES hosts are consistent with the observed SSMF. Both ends of the SSMF are well-fitted, including two distinct features -- the shelf around $M_\star \sim 10^{10}\,M_\odot$ and the flattening below $M_\star \sim 10^{6.5}\,M_\odot$.

\begin{table*}[]{}
\setlength{\tabcolsep}{9pt}
\renewcommand{\arraystretch}{1.2}
\caption{Satellite–Subhalo Connection Model Constraints Derived using the ELVES Satellite Galaxies}
\centering 
\begin{tabular}{c c c c} \hline\hline 
Model Parameter & Parameter Description & 68\% Credible Interval & 95\% Credible Interval \\ [0.5ex]
\hline 
$\alpha$ & The faint-end slope of the $M_\mathrm{peak}-M_\star$ relation & $2.093 < \alpha < 2.115$ & $2.083 < \alpha < 2.125$ \\
\hline 
$\sigma$ & The scatter & $0.019 < \sigma < 0.132$ & $ 0.003 < \sigma < 0.205$ \\
\hline 
\hline 
\end{tabular}
\label{table:results}
\end{table*}

\subsection{A Growing Scatter Model} \label{sec:growing_scatter}

Previous studies of the SHMR at high masses \citep{Moster:2010, Moster2013, Behroozi:2010, Behroozi2013, Behroozi:2019}, as well as much of the work on the Milky-Way satellite SHMR \citep{GK2017:SMHM, Nadler:2019, Nadler:2020a}, considered a ``constant scatter'' model where a symmetric, log-normal scatter $\sigma$ in $M_\star$ at fixed $M_\mathrm{peak}$ is assumed. However, recent results from hydrodynamical simulations (e.g. \citealt{Sawala2016:scatter}; \citealt{Munshi2021:SMHM}) as well as from simple halo occupation models \citep{Smercina:2018} motivated the consideration of a growing scatter model in which the scatter grows linearly with decreasing log peak halo mass for $M_\mathrm{peak} \le M_1$, at the dwarf regime.

Therefore, we use our data and methodology to also examine a satellite–subhalo connection model with a growing scatter. We repeat the same procedure described in Section \ref{sec:connection}, using the same model to describe the mean SHMR relation (Equation \ref{eq:model_median_mstar}) but following \citet{GK2017:SMHM}, we let the scatter grow as a function of $M_\mathrm{peak}$:

\begin{equation} \label{eq:scatter}
\sigma (M_\mathrm{peak}) = \sigma_0 + \nu(\log_{10}M_{\mathrm{peak}} - \log_{10} M_1)
\end{equation}

\noindent where $\sigma$ is the scatter at $M_\mathrm{peak}$, $\nu$ is the rate at which the scatter changes, and $\sigma_0$ is the scatter at and above $M_\mathrm{peak}=M_1$.

\begin{figure*}
\centering     
\subfigure{\label{fig:a}\includegraphics[width=85mm]{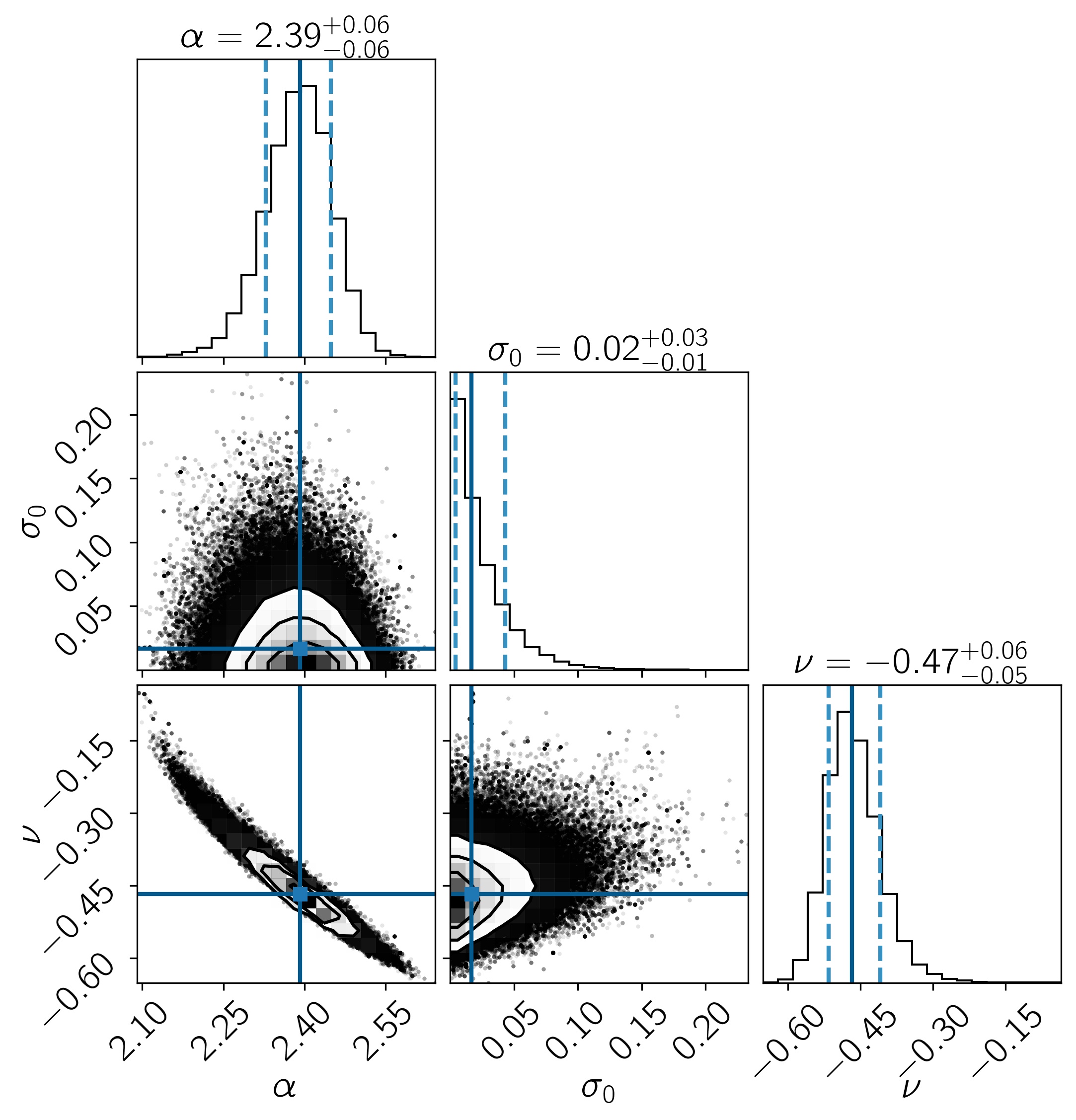}}
\subfigure{\label{fig:b}\includegraphics[width=85mm]{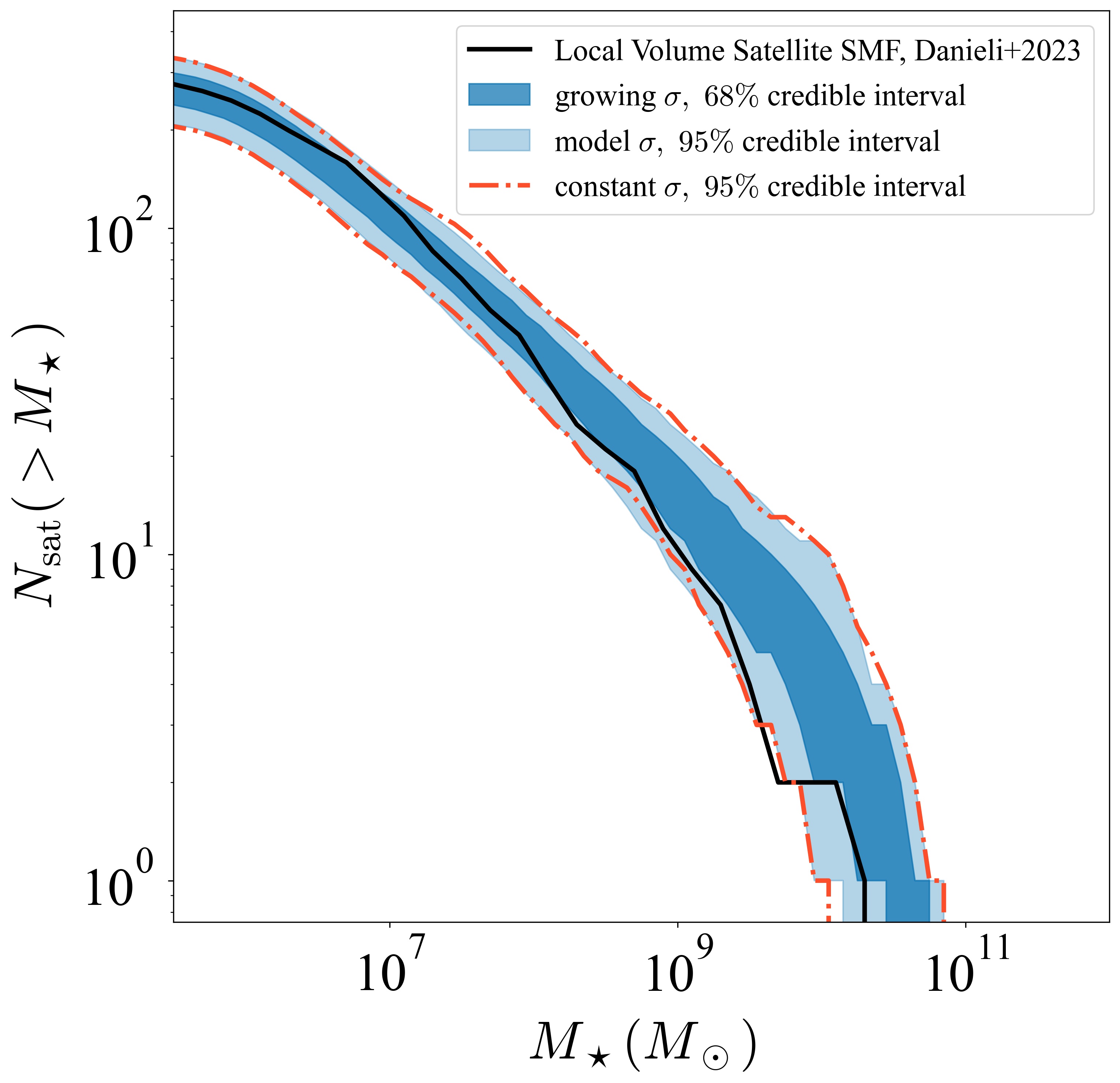}}
\caption{Results for the growing scatter satellite-subhalo connection model. \textit{Left panel:} posterior distribution of the parameters in the growing scatter model for the satellite SHMR. Solid blue lines show the median (50\%) and dashed light blue lines show the 68\% confidence interval. \textit{Right panel:} the 68\% (blue) and 95\% (light blue) confidence intervals for the satellite stellar mass function predicted by the growing scatter model, compared to the satellite stellar mass function observed with ELVES (black). The predictions are consistent with the observed SMF and compared to our fiducial model with a constant scatter.}
\label{fig:growing_scatter}
\end{figure*}

Similarly to the constant scatter model parameter inference, we run $5 \times 10^4$ MCMC iterations using 20 walkers and discard the first 5000 steps. The results are shown in Figure \ref{fig:growing_scatter}. The left panel shows the posterior distribution for the growing scatter model parameters. Compared to the constant scatter model, the slope is steeper with $\alpha_\mathrm{grow}=2.39 \pm 0.06$ ($\alpha_\mathrm{const}=2.10 \pm 0.01$), and the scatter grows at a rapid rate with $\sigma_0 = 0.02^{+0.03}_{-0.01}$ and $\nu = -0.47^{+0.06}_{-0.05}$. The right panel of Figure \ref{fig:growing_scatter} compares the observed SSMF with the $68\%$ (blue) and $95\%$ (light blue) credible intervals for the stellar mass function, drawn from the posterior distribution of the growing scatter model. The observed SSMF is well reproduced as in the case of the constant scatter model. We discuss these results in Section \ref{sec:discuss_results}.

\subsection{The Satellite Stellar to Halo Mass Relation}\label{sec:results_shmr}

\begin{figure*}[t]
 \centering
 \includegraphics[width=1.0\textwidth]{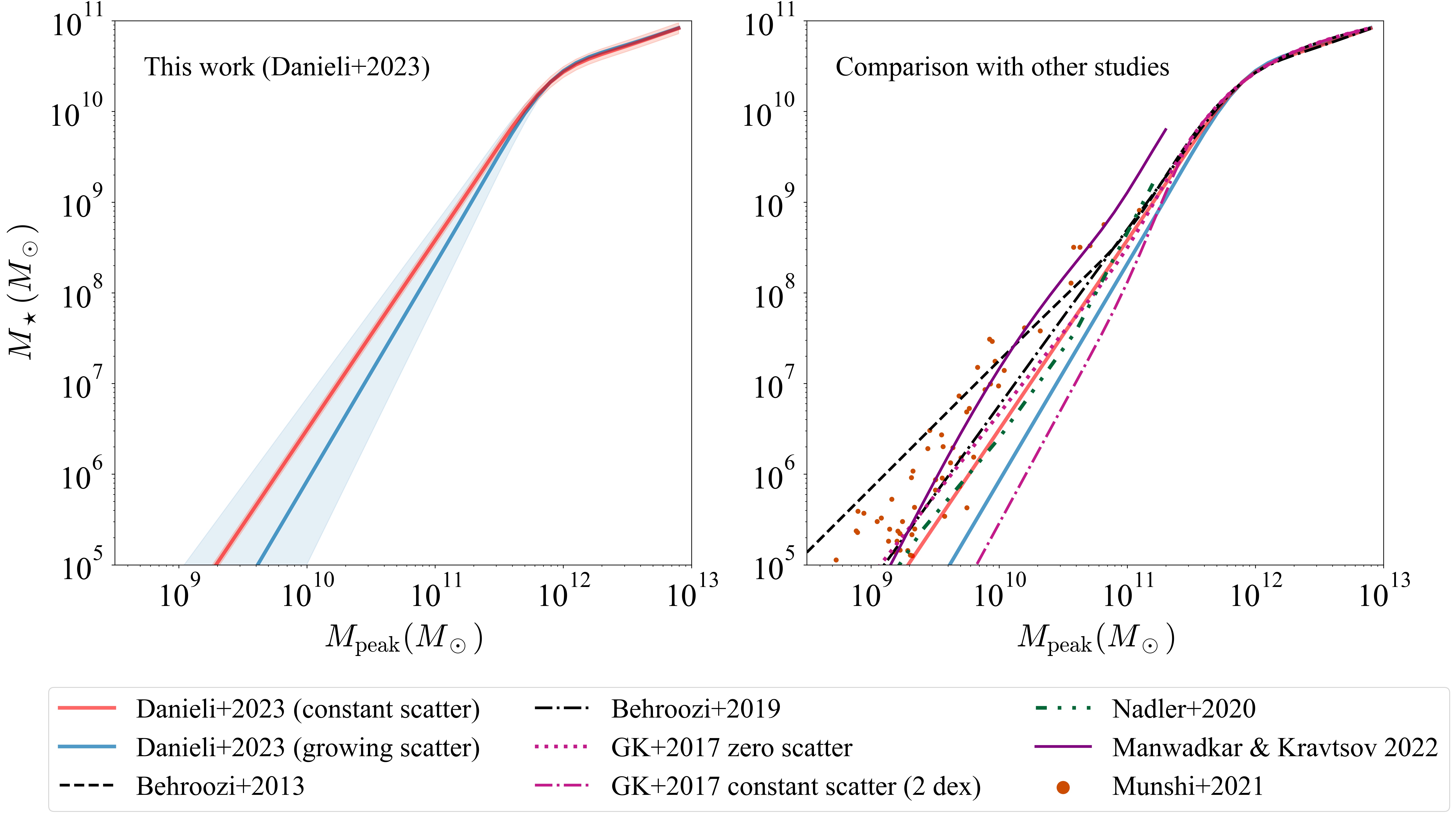}
  \caption{\textit{Left}: the Satellite Stellar to Halo Mass Relations obtained in this work: the red curve shows the median relation for our fiducial model described by two parameters and a constant scatter shown in the shaded region (\S \ref{sec:results_fiducial}). The blue curve shows the median relation for the growing scatter model described by three parameters (\S \ref{sec:growing_scatter}). \textit{Right}: the same relations compared to recent predictions from previous studies. \citet{GK2017:SMHM} and \citet{Nadler:2020a} present models obtained using the Milky-Way satellites, and the \citet{Munshi2021:SMHM} galaxies are the result of high-resolution hydrodynamical simulations.}
  \label{fig:shmr}
\end{figure*}

We show the constrained satellite stellar to halo mass relation for the fiducial model (constant scatter model; \S \ref{sec:results_fiducial}), and the growing scatter model (\S \ref{sec:growing_scatter}) in Figure \ref{fig:shmr}. In both cases, we adopt the median of the posterior distribution as a point estimate used in describing the functional form of the SHMR. The left panel shows the two relations obtained in this work. In red, we show the inferred constant scatter model, and in blue, we show the inferred SHMR assuming a growing scatter. The two models are different from each other in both the constrained slope and the scatter. The fiducial, constant scatter model favors a moderately shallower slope ($\alpha_\mathrm{const}=2.10 \pm 0.01$) and a very low scatter ($\sigma_\mathrm{const}$ = 0.06) where the growing scatter model is constrained to have a steeper slope ($\alpha_\mathrm{grow}=2.39 \pm 0.06$) and a scatter that grows rather quickly below $\log_{10}M_\mathrm{peak} \sim 11.9\,M_\odot$. The growing scatter reaches a value of 0.44 dex at $M_\mathrm{peak}=10^{11}\,M_\odot$, 0.9 dex at $M_\mathrm{peak}=10^{10}\,M_\odot$, and 1.4 dex at $M_\mathrm{peak}=10^{9}\,M_\odot$. This trade-off between a shallower slope combined with a small scatter and a steeper slope combined with a larger scatter has been discussed in \citet{GK2017:SMHM}. Both models produce similar SSMFs which are also consistent with the observed Local Volume SSMF (Figures \ref{fig:constant_scatter} and \ref{fig:growing_scatter}). When the scatter is allowed to grow, the slope, $\alpha$, must also grow (become steeper) to avoid overproducing the observed SSMF.

In the right panel of Figure \ref{fig:shmr}, we compare the two relations with other constrained SHMR relations from the literature. Our fiducial, constant scatter model, shown in red, is consistent with previous relations derived using the Milky Way satellite population. In particular, it is very similar to the relation presented in \citet{Nadler:2020a}. It is also partly in agreement, though slightly steeper, with the ``zero scatter'' relation presented in \citet{GK2017:SMHM} and the extrapolation of the \citet{Behroozi:2019} median SHMR presented for satellite galaxies at $z=0$, which has the functional form we adopted in this work. As noted, the growing scatter model (blue curve) presents a steeper slope, similar to the 2 dex constant scatter model from \citet{GK2017:SMHM}. We also compare our derived SHMR to results from hydrodynamical simulations. The brown dots show the simulated galaxies from a suite of high-resolution zoomed-in hydrodynamical simulations (the MARVEL-ous Dwarfs and Justice League; \citealt{Munshi2021:SMHM}). Similar to our growing scatter model, the scatter increases with decreasing $M_\mathrm{peak}$ and spans a similar range in stellar mass for a fixed $M_\mathrm{peak}$. However, the galaxies simulated in \citet{Munshi2021:SMHM} are generally offset to higher stellar masses at a fixed $M_\mathrm{peak}$ compared to our growing scatter model.

\subsection{Host-to-Host Scatter}\label{sec:host_to_host_scatter}

\citet{Mao:2021} reports a significant correlation between the host $M_{K_s}$ magnitude and the total satellite count, for satellites brighter than $M_{r}=-12.3\,\mathrm{mag}$ within $150\,\mathrm{kpc}$ of the host for galaxy groups studied in the SAGA survey. As the groups surveyed in ELVES span approximately an order of magnitude in the central galaxy stellar mass ($\log M_\star = 9.9 - 11.1\,M_\odot$), we expect to be able to detect a similar correlation. \citet{Carlsten:2021c} examine this host-to-host scatter for a subset of the ELVES groups, identifying a strong correlation between satellite abundance and host mass. We revisit this measurement using the sample of 27 ELVES hosts and compare it with what our new satellite SHMR model gives. 

\begin{figure}
 \includegraphics[width=0.49\textwidth]{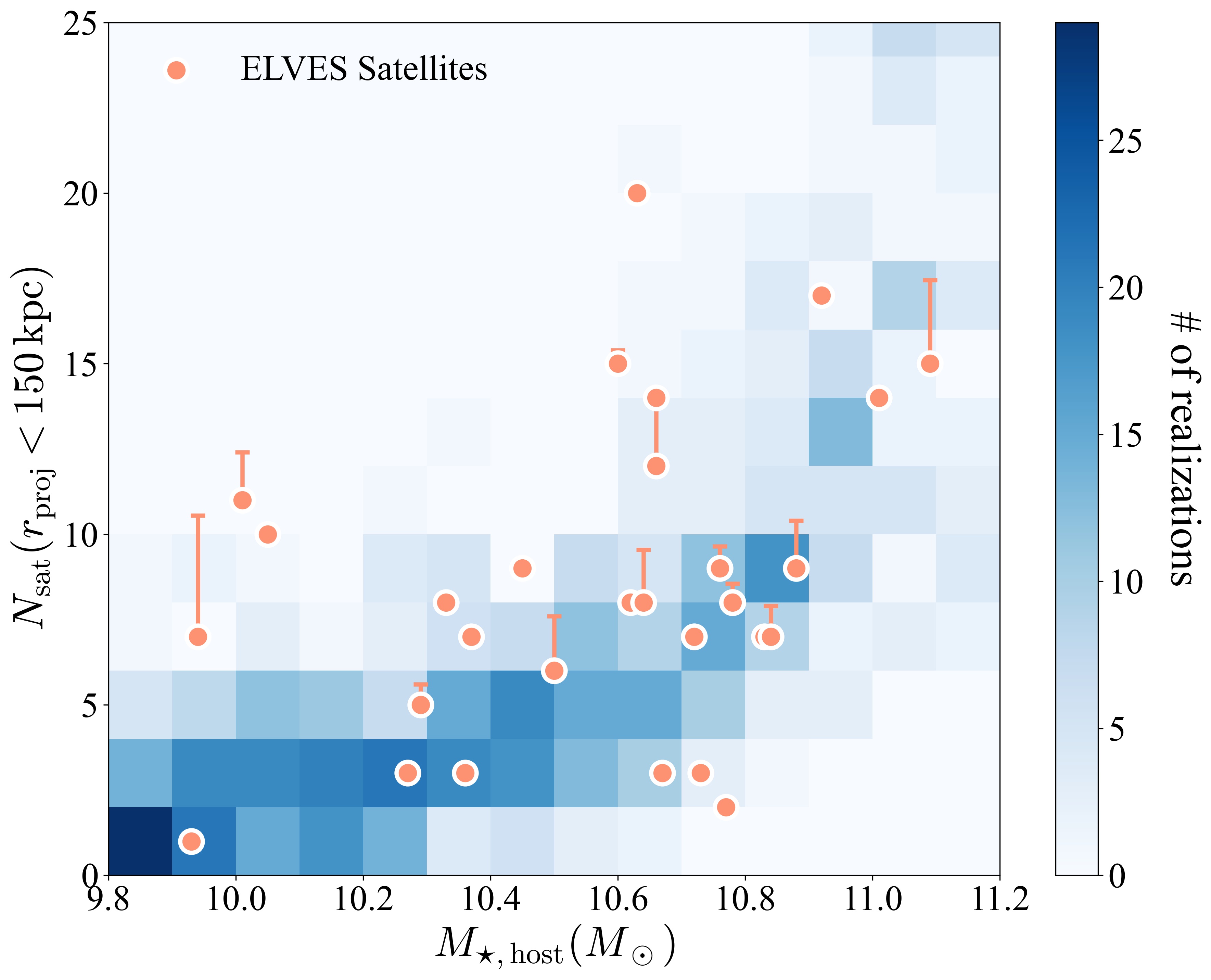}
  \caption{The number of satellites, $N_\mathrm{sat}$, versus the stellar mass of the host inferred from its $M_K$-band group magnitude. Orange circles show the observed satellites of the 27 ELVES groups considered in this work, where satellites within $r_\mathrm{proj}=150\,\mathrm{kpc}$ are included, and upper error bars represent the uncertainty in $N_\mathrm{sat}$ due to unconfirmed candidates (Section \ref{sec:elves}). The blue-shaded color map shows the number of realizations in $N_\mathrm{sat}$ and $M_{\star,\mathrm{host}}$ bins from simulating the satellite populations hosts galaxies (50 realizations pere $M_{\star, \mathrm{host}}$), by populating their predicted subhalos (from \satgen) with our fiducial satellite SHMR. The observed and simulated trends are largely consistent with each other, though in some cases a large spread in $N_\mathrm{sat}$ for a given $M_\star$ suggests that the number of satellites might serve as a useful tracer for the physical processes that take place in these groups.}
\label{fig:nsat}
\end{figure}

To calculate the theoretical satellite richness, we do the following: for a given stellar mass, $M_\star$, we calculate its halo mass using the \citet{RP:2017} SHMR at $z=0$, assuming a constant scatter of $0.15\,\mathrm{dex}$. Then, we get its subhalo population from the \satgen library described in Section \ref{sec:satgen}. The population of satellite galaxies is then forward-modeled as described in Section \ref{sec:modelsmf}, using the fiducial SHMR (\S \ref{sec:results_fiducial}), where the observational selection functions and uncertainties are applied. This procedure is repeated 50 times for each stellar mass. We show the results in Figure \ref{fig:nsat}. For both the ELVES satellites and the forward-modeled satellites, we only include galaxies within a projected distance from the host of $150\,\mathrm{kpc}$, as all ELVES hosts are surveyed to at least this projected distance from the host center. The overall agreement between the observed satellites (orange circles) and the simulated satellites (blue-shades background color map) indicates that the model describes well the correlation between the host stellar mass and the satellite abundance. While in general, it is apparent that such correlation exists, the groups in several narrow host mass bins, such as e.g. $\log M_\star = 10-10.2$ span a considerable range in the number of satellites, ranging from just one satellite to $11^{+1.4}_{-0.0}$ satellites within $r_\mathrm{proj}=150\,\mathrm{kpc}$. This large spread in the number of satellites for a fixed host stellar mass suggests that the satellite population can be significantly impacted by the specific formation and evolution history of a specific group. We discuss this aspect further in Section \ref{sec:discussion}. Finally, we note that increasing the sample of surveyed groups down to lower-mass hosts (e.g. \citealt{Roberts:2021}) would provide useful information about the $M_\star - N_\mathrm{sat}$ relation.

\section{Discussion and Summary} \label{sec:discussion}

In this paper, we present a model for the connection between satellite galaxies that orbit central galaxies in Local Volume groups and their host dark matter subhalos. The fiducial model we constrain assumes a median relation of the form $M_\star \approx M_\mathrm{peak}^\alpha$ at the low mass end ($\log M_\mathrm{peak} \lesssim \log M_1=11.9\,M_\odot$), with a constant scatter for all $M_\mathrm{peak}$  (Equation \ref{eq:model_median_mstar}). To constrain the model parameters, we used a new, unique sample of 250 confirmed and 71 candidate satellite galaxies around 27 host galaxies in the Local Volume ($D \lesssim 12\,\mathrm{Mpc}$), constructed as part of the Exploration of Local VolumE Satellites (ELVES) Survey \citep{Carlsten:2022}. We infer the model parameters' posteriors by performing a Bayesian comparison between the observed Local Volume satellite stellar mass function and a forward-modeled stellar mass function, by populating dark matter subhalos generated with the semi-analytic model \satgen, and carefully accounting for observational biases and selection effects. In what follows we discuss key results from this work and review caveats and future work.

\subsection{Moderate Slope, Small Scatter}\label{sec:discuss_results}

Our fiducial model results in a well-constrained, moderate slope of $\alpha_\mathrm{const}=2.10 \pm 0.01$, steeper than relations reported for higher mass galaxies ($\alpha=1.6 - 1.9$; \citealt{Behroozi2013, Moster2013, GK2017:SMHM}), and shallower than the slope presented in \citet{GK2017:SMHM} ($\alpha=2.6$) using the MW and M31 satellites and assuming a $2\,\mathrm{dex}$ scatter at all subhalo masses. We get consistent results with those presented in \citet{Nadler:2020a}, derived statistically from fitting the MW satellite population alone. The constraints on the faint-end slope are also consistent with that presented in \citet{Behroozi:2019} for satellite galaxies at $z=0$ ($\alpha_\mathrm{sat}=1.959^{+0.172}_{-0.022}$), though it utilized a higher mass sample than is used here and in \citet{Nadler:2020a}. Our 2-parameter fiducial model requires a very small scatter with a posterior median of $0.06^{+0.07}_{-0.05}$ and a three-sigma credible interval of $\sigma_\mathrm{const}<0.3$.

These values are consistent with the scatter inferred from analyses for more massive systems and consistent with statistical constraints using the full current census of MW satellite \citep{Nadler:2020a}. \citet{Kravtsov:2022} and \citet{Manwadkar:2022} modeled the formation of dwarf galaxies in MW-analogs using a ``regular'' or ``bathtub''-type model (GRUMPY; Galaxy formation with
RegUlator Model in PYthon). They show that different GRUMPY models result in different $M_\star - M_\mathrm{peak}$ relation slopes, depending on the wind mass loading factor model parameters. However, in all models (Figures 3, 4, and 5 in \citealt{Kravtsov:2022}), the scatter stays small down to $M_\mathrm{peak}\sim 10^{10} M_\odot$. Their fiducial model with $\eta_\mathrm{norm}=3.5$ (the normalization of the mass loading factor) and $\eta_\mathrm{p} = 0.35-0.4$ (the power-law exponent in the power law dependence of mass loading factor on $M_\star$) result in a very small scatter ($<0.1\,\mathrm{dex}$) for $M_\mathrm{peak}\gtrsim 10^{10}M_\odot$. In the ultra-faint dwarf galaxy regime, below our completeness limit, the scatter reaches $\sim 0.3-0.4$ dex. The low scatter in this mass range is also typical across a range of simulations implemented with different physics (\citealt{Applebaum:2021, Orkney:2021, Gutcke:2022, Hopkins:2023}; also see a summary figure in Appendix B of \citealt{Manwadkar:2022}).

We also examine whether a growing scatter model (see Equation \ref{eq:scatter} for the scatter's functional form in this case) can provide a good fit to the observed satellite stellar mass function. Modeling the $M_\star - M_\mathrm{peak}$ relation with a mass-dependent growing scatter model is physically motivated by recent results from hydrodynamical simulations -- low-mass galaxies are more sensitive compared to their high-mass counterparts to a variety of baryonic and environmental processes \citep{Sawala2016:scatter, Rey:2019, Munshi2021:SMHM}. The growing scatter model is constrained to have a steeper slope ($\alpha_\mathrm{grow} = 2.39 \pm 0.06$) and the scatter increases rapidly for decreasing subhalo mass, $M_\mathrm{peak}$ (see Figure \ref{fig:shmr}). We find that subhalos with a peak mass of $\sim 10^{11}\,M_\odot$ span $\sim 0.5\,\mathrm{dex}$ in $\log\,M_\star$ and subhalos with $M_\mathrm{peak} \sim 10^9 - 10^{10}\,M_\odot$ can host galaxies across $\sim 1-1.5\,\mathrm{dex}$ in stellar mass.
Similarly to \citet{GK2017:SMHM}, we see a degeneracy between the SHMR slope and the scatter in $\log M_\star$ at a given $M_\mathrm{peak}$. This degeneracy is indicated in Figure \ref{fig:growing_scatter} by the strong anti-correlation between the slope $\alpha$ and $\nu$, the rate at which the scatter changes -- $\nu$ decreases (i.e. the scatter grows) as the slope increases. A shallow slope combined with a large scatter will over-predict the number of satellites across the entire stellar mass range compared to the observed satellite stellar mass function.

\subsection{Is There a Universal Satellite Stellar to Halo Mass Relation at the Dwarf Regime?}\label{sec:universal}

Compared to earlier studies which primarily used the MW and M31 satellite systems to study to SHMR below $M_\mathrm{peak} \sim 10^{11}\,M_\odot$, here we include a substantially larger sample of satellite galaxies, from a variety of groups in the Local Volume. Although each group has a distinct evolutionary history, we find a remarkable consistency in the stacked Local Volume SHMR's parameterized abundance matching analysis, inferring a satellite SHMR that aligns closely with the relation inferred solely from the MW satellite system. Our data suggest that the satellite SHMR has a universal functional form across hosts. The small scatter in the stellar mass ($M_\star$) at peak halo mass ($M_\mathrm{peak}$) strongly supports the idea of a universal SHMR. In order for the relation to be so independent of detailed accretion histories, galaxy formation must be extremely well-regulated, even down to halo peak masses of $M_\mathrm{peak}\sim 10^9-10^{10}\,M_\odot$ (corresponding to stellar masses of $M_\star \sim 10^6\,M_\odot$). That is, on average, galaxies occupy dark matter halos at $M_\mathrm{peak}$ in a very predictable manner, similar to massive galaxies. It is important to acknowledge that our use of peak halo masses implies that the SHMR at z=0 may exhibit a larger scatter in the relation since subhalos evolve once they become part of a group.

An intriguing thought arises regarding whether the practice of stacking satellite galaxies from diverse groups, aimed at increasing the statistical power, leads to the loss of scatter-related information. It is analogous to the method employed in constraining the SHMR for large samples of field galaxies at higher masses, where the stellar mass function is composed of numerous galaxies of mixed large-scale environments and evolutionary histories. The ELVES dataset, encompassing 27 groups, may have adequate constraint capabilities to study this scatter either by simultaneously fitting all groups individually (with a common slope) or by fitting each group separately. However, such analyses are deferred to future investigations, potentially leveraging even larger samples.

Nevertheless, the results presented in this study showcase that the average relation between $M_\star$ and $M_\mathrm{peak}$ successfully reproduces at least two observables on an individual group level. First, as shown in Figure \ref{fig:nsat} and discussed in Section \ref{sec:host_to_host_scatter} -- both the correlation and the scatter between the host stellar mass ($M_{\star, \mathrm{host}}$) and the number of satellites ($N_\mathrm{sat}$) is well reproduced when we populate dark matter subhalos with galaxies using the fiducial SHMR constrained using the stacked Local Volume satellite stellar mass function. The scatter in $N_\mathrm{sat}$ is a direct result of the scatter in the number of dark matter subhalos produced by \satgen for a fixed $M_\mathrm{vir}$ (this is also demonstrated in Figure 2 in \citealt{Jiang2021:SatGen}).

\begin{figure*}[t]
 \centering
 \includegraphics[width=0.82\textwidth]{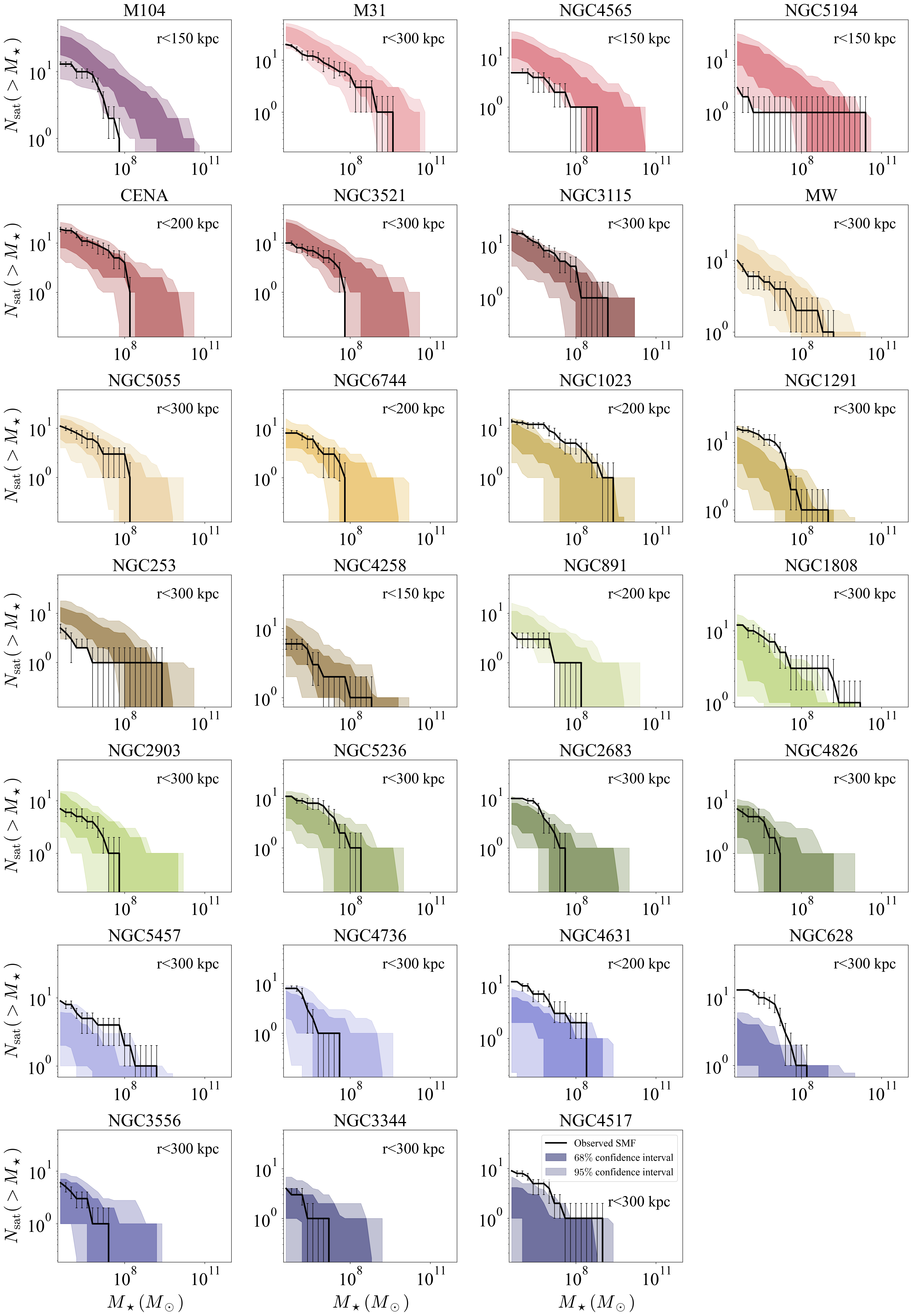}
  \caption{Posterior predictive plots for all 27 ELVES hosts from the stacked Local Volume satellite inference, ranked order by their group $K$-band magnitude from top left to bottom right. Similarly to Figure \ref{fig:constant_scatter}, the black curves show the satellite stellar mass function observed with ELVES for each one of the hosts and the colored shaded curves show the $68\%$ and $95\%$ credible intervals for the predicted satellite stellar mass function, using our constrained fiducial model (\ref{sec:results_fiducial}) and 50 \satgen realizations for each host. The model describes the data well in the vast majority of cases.}
\label{fig:single_host_smf}
\end{figure*}

Second, in Figure \ref{fig:single_host_smf}, we show the predicted satellite stellar mass function for each host, compared to the observed stellar mass functions for individual groups as observed in ELVES. We generate 50 realizations of subhalo populations for each host and populate them with galaxies using our fiducial (constant scatter) SHMR. The shaded regions show the $68\%$ and $95\%$ credible intervals where the black curves show the observed stellar mass functions, with error bars representing one standard deviation, bootstrapped from the observed SMF. The predicted satellite stellar mass functions for individual hosts are in agreement with the observed SMFs, within the errors, for the vast majority of hosts. This implies that we are able to predict the population of satellite galaxies in MW-like groups, with high certainty, using the SHMR derived in this work by stacking the satellite galaxies from the 27 Local Volume hosts. In one case (NGC628), the well-sampled SMF is not predicted by the model. This host is one of the richest hosts in the ELVES sample in terms of its number of satellites ($N_\mathrm{sat}=13$) for its host stellar mass ($\log (M_{\star, \mathrm{host}}/M_\odot)=10.45$). It would be interesting to investigate what might be causing this disagreement (e.g. an unusual growth history, etc.).

\subsection{Caveats and Future Work}

We briefly discuss the limitations and exciting possible avenues for future follow-up work.

\noindent
\textbf{Model Assumptions and Choices:} We use the \satgen semi-analytical model to generate the dark matter subhalos we populate with the observed ELVES satellites. However, \satgen provides a particular model for the evolution and disruption of the subhalos and using a different model can potentially change the results. It will be interesting to see what happens when we use different suites of simulations that differ in their baryonic prescriptions. 

It would also be interesting to test the impact of various choices made in the \satgen run on the derived SHMR. Two particular examples of such choices are the adopted halo response model and having the host halo include a disk (see Section \ref{sec:satgen}). Both choices will change the overall number of subhalos above some peak mass, $N(>M_\mathrm{peak})$ which will directly affect the number of satellites that can be hosted by a subhalo at that peak mass. \citet{Jiang2021:SatGen} showed that the halo response has a small effect on the subhalo mass function while including a baryonic disk reduces the abundance of surviving satellites within 300 kpc by $\sim 20\%$. In this work, when generating the library of \satgen halos, we assumed an embedded galactic disk that comprises $10\%$ of the mass of the halo, for all host halos. However, ELVES includes a variety of early and late-type hosts, some of which do not include a disk. It will be interesting to quantify how the disk affects the posterior Satellite SHMR.

\noindent
\textbf{Future Observational Work:} Expanding the systematically surveyed Local Volume satellite dwarf galaxies from five to a total of 30, as done in ELVES \citep{Carlsten:2019b, Carlsten:2020, Carlsten:2021a, Carlsten:2021b, Carlsten:2021c, Carlsten:2022}, provides a unique perspective on dwarf satellite galaxies in a way that is not limited to the particular case of the MW (see also \citealt{Geha:2017, Mao:2021} for the SAGA Survey). In order to adequately sample the many environments and evolutionary histories that dwarf galaxies experience, the surveyed volume should be increased even more. Additionally, it would be intriguing to survey the satellite populations orbiting smaller mass hosts, as ELVES lacks such Local Volume hosts. Likewise, deriving the SHMR for low-mass galaxies using a large volume-limited sample of ``field'' (isolated) dwarf galaxies would be extremely useful. Comparing the satellite and field SHMR is essential for establishing how much each physical process that affects those measurements contributes \citep{Watson:2013, Read:2017, Engler:2021a}. For example, pre-processing of dwarfs during infall and within the intragroup and intracluster media can dramatically change their properties \citep{Putman:2021}. Constraints on the Satellite SHMR can also be improved by surveys going deeper. In particular, the low mass end of the stellar mass function holds a lot of the constraining power, mostly on the scatter, and this is also the part that suffers the most from the observational uncertainties (\S \ref{sec:survey_charc}). So while we fold these observational selection functions into the modeling, minimizing their uncertainties will surely help.

\subsection{Summary}
Our inferred model parameters and especially the scatter provide important constraints on galaxy formation, dark matter models, and the interplay between the two. The choices made in hydrodynamical simulations, e.g. the dark matter particle type, the cold gas fractions, star formation, and feedback prescriptions, would need to be able to reproduce the $M_\star - M_\mathrm{peak}$ relation presented in this work. Combined with the additional constraints on the observed properties of dwarf galaxies (e.g. their distribution in the size-mass-surface brightness plane, the measured dark matter density profile of individual galaxies, etc.), would help break degeneracies between these various components. Importantly, the vast majority of satellite-subhalo connection studies thus far were informed by the satellite population of the MW. In this paper, we use the satellite population of 27 Local Volume groups and this nearly volume-limited sample of galaxy groups is expected to sample better the true distribution of groups. Since simulations also often use a suite of realizations, this approach would provide a less biased tracer of the satellite population around such hosts and, as a result, of the $M_\star - M_\mathrm{peak}$ relation for satellite galaxies.

\acknowledgments

We thank Andrey Kravtsov, Risa Wechsler, Frank van den Bosch, Imad Pasha, Ethan Nadler, Peter Melchior, Jiaxuan Li, and Tjistske Starkenburg for very useful discussions related to this work.
S.D. is supported by NASA through Hubble Fellowship grant HST-HF2-51454.001-A awarded by the Space Telescope Science Institute, which is operated by the Association of Universities for Research in Astronomy, Incorporated, under NASA contract NAS5-26555. J.E.G. acknowledges support from NSF Astronomy and Astrophysics Grant \#1007052.

\software{Astropy \citep{Astropy-Collaboration:2013aa, Astropy-Collaboration:2018, Astropy-Collaboration:2022}, 
  matplotlib \citep{Hunter:2007aa}, 
  numpy \citep{Van-der-Walt:2011aa}, 
  scipy \citep{Virtanen:2020},
  corner \citep{Foreman-Mackey:2016}
.}

\bibliographystyle{aasjournal}

\end{document}